\documentstyle[epsfig]{mn}
\font\japit = cmti10 at 10truept
\title[Eulerian bias and the galaxy density field]
{\vglue-3.0truecm
\centerline{\japit Accepted for publication in Monthly Notices of the R.A.S.}
\vglue 2.5truecm\noindent
Eulerian bias and the galaxy density field}

\author[R.G. Mann, J.A. Peacock and A.F. Heavens]
{R.G. Mann,$^{1,2,3}$ J.A. Peacock$^{4}$ and A.F. Heavens$^1$\\
$^1$Institute for Astronomy, Department of Physics and Astronomy, University 
of Edinburgh, Blackford Hill, Edinburgh EH9 3HJ\\
$^2$Astronomy Unit, 
Queen Mary and Westfield College, Mile End Road, London E1 4NS\\
$^3$Present Address: Astrophysics Group, Blackett Laboratory, 
Imperial College, Prince Consort Road, London SW7 2BZ\\
$^4$Royal Observatory, Blackford Hill, Edinburgh, EH9 3HJ}

\newcommand{\beq}{\begin{equation}}
\newcommand{\eeq}{\end{equation}}
\newcommand{\bfi}{\begin{figure}}
\newcommand{\efi}{\end{figure}}
\newcommand{\bit}{\begin{itemize}}
\newcommand{\eit}{\end{itemize}}
\newcommand{\rmd}{{\rm d}}
\newcommand{\myref}[1]{\noindent \hangindent=0.5in \hangafter=1 #1 \par}

\renewcommand{\baselinestretch}{1.0}

\def\kms{\;{\rm km\,s^{-1}}}
\def\kmsmpc{\;{\rm km\,s^{-1}\,Mpc^{-1}}}
\def\hompc{\;h\,{\rm Mpc}^{-1}}
\def\mpcoh{\;h^{-1}\,{\rm Mpc}}
\def\ss{\scriptscriptstyle\rm}

\begin{document}

\maketitle
\renewcommand{\baselinestretch}{1.0}
\begin{abstract}
\noindent
We investigate the effects on cosmological clustering statistics of
empirical biasing, where the galaxy distribution is a local 
transformation of the present-day Eulerian density field.
The effects of the suppression of galaxy numbers in voids, and 
their enhancement in regions of high density, are considered, 
independently and in combination.
We compare results from numerical simulations with 
the predictions of simple analytic models.
We find that the bias is generally scale-dependent, so that the shape
of the galaxy power spectrum differs from that of the underlying mass
distribution.  The degree of bias is always a monotonic function of
scale, tending to an asymptotic value on scales where
the density fluctuations are linear.  The scale
dependence is often rather weak, with many reasonable
prescriptions giving a bias which is nearly independent of scale.

We have investigated whether such an Eulerian bias can reconcile a range of 
theoretical power spectra with the twin requirements of fitting the 
galaxy power spectrum and reproducing the observed mass-to-light ratios in
clusters.   It is not possible to satisfy these constraints for 
any member of the family of CDM-like power spectra in an Einstein - de Sitter 
universe when normalised to match {\em COBE\/} on large scales
and galaxy cluster abundances on intermediate scales.
We discuss what modifications of the mass power spectrum might produce
agreement with the observational data.

\end{abstract} 
\section{INTRODUCTION}

The great advances in observational cosmology during the past decade have
provided a wealth of new data on galaxy clustering. These results
facilitate a phenomenological approach to the study of
large-scale structure, in which the mass power
spectrum is reconstructed directly from the observed clustering statistics
(e.g. Peacock \& Dodds 1994, PD94). This method is
complementary to the hitherto conventional technique of judging a pet
cosmogony on the basis of its predictions for a series of 
observationally-determined quantities, and has some advantages so long as
no specific model appears
capable of accounting for the full set of observational data.

There is, however, a fundamental problem with both the empirical and the
a priori approaches
to large-scale structure. The linear mass power
spectrum that one deduces from the observed clustering depends on, and 
may be very sensitive to, an assumed
relationship between the distributions of mass and galaxies in the
Universe, and ignorance of the detailed processes through which galaxies
form leaves this relationship poorly constrained.
Different classes of galaxy are observed to cluster differently (see,
e.g., the compilation by PD94), implying that
at least some galaxies are biased and do not directly trace
the mass distribution. Bias is also required by advocates of an
Einstein -- de Sitter universe, to reconcile observed cluster $M/L$
values that imply $\Omega \simeq 0.2$  with the
assumed critical density. 
In the absence of a complete understanding of galaxy 
formation and evolution,
large-scale structure theorists are left to model bias in
as plausible a manner as they can.

Many such attempts have used the high-peak bias method (Davis et al. 
1985; Bardeen et al. 1986, BBKS).
We will term this a {\it Lagrangian\/} bias scheme, since it 
identifies the sites of 
nascent objects (galaxies or clusters) with peaks in the {\it initial\/} density 
field, smoothed on a scale appropriate to the mass of the objects
under consideration. 
If the initial density field is Gaussian, then these peaks are more
strongly clustered that the mass distribution as a whole, producing the desired
bias. {\em N\/}--body simulations (e.g. Katz, Quinn \& Gelb
1993) have revealed a poor correspondence between the
particles found in galaxy halos selected at late times and those
located at appropriate peaks in the initial density field, thus  
casting doubt on Lagrangian galaxy bias prescriptions (although the
work of Mo \& White 1996 and Mo, Jing \& White 1996 suggests
Lagrangian bias may work statistically):
this is not surprising, since the density field
smoothed on the scale of galaxies is highly non-linear. For
cluster-sized halos, however, Lagrangian bias may still be appropriate
(Kaiser 1984; Cole \& Kaiser 1989; Mann, Heavens \& Peacock 1993), since their spatial
distribution has undergone far less dynamical evolution.

In reality, galaxy formation must be more complex than this,
involving a range of feedback mechanisms (Dekel \& Rees 1987; Babul
\& White 1991).
One might hope that all these issues 
would eventually be clarified by large
numerical simulations which include a hydrodynamical treatment of 
baryons, as well as dissipationless dark matter, but such codes are
in their infancy. 
An intriguing early result from one such code was the observation by Cen \&
Ostriker (1992; CO) of a tight correlation between the present-day (Eulerian) density
field, $\rho_{\rm m}$, and the local number density, $\rho_{\rm g}$, 
of galaxies, formed using what those
authors termed a `heuristic but plausible' prescription, which 
creates a dissipationless proto-galactic particle wherever the local
baryonic component is sufficiently dense, rapidly cooling, contracting
and Jeans unstable. Together with the observed uniformity of $M/L$
values as a function of scale,
these results motivate the general idea of {\it Eulerian\/}  bias 
models, in which the galaxy number density  
at redshift zero is a function of the local value of the evolved
mass density field. Some general aspects of such prescriptions were 
considered
by Coles (1993), who deduced various important inequalities relating the 
mass and galaxy correlation functions in these models.

Our approach here is as follows. We make no assumptions 
concerning the
physical processes that lead to bias, simply taking  
their overall effect to be
a local transformation of the Eulerian density field,  the principal
features of which are the suppression of galaxy numbers in 
low-density regions
and their enhancement in regions of high density. In Section 2 we
outline the set  of bias prescriptions combining these
components which are to be studied here, using the combination of
numerical and analytical methods described in Section 3, while Section
4 shows the effect of biasing CDM-like mass models by these methods.
We focus on the question of the scale-dependence of the Fourier space
bias parameter, $b(k)$, defined
as the square root of the ratio of the galaxy and mass power
spectra:
\beq
b(k)\equiv [\Delta_{\rm g}^2(k)/\Delta_{\rm m}^2(k)]^{1/2},
\eeq
where $\Delta_{\rm g}^2(k)$ and $\Delta_{\rm m}^2(k)$ are, respectively,
the galaxy and mass power spectra. We employ the
dimensionless form of the power spectrum, $\Delta^2(k)$, which is
defined to be the variance per $\ln k$ [i.e. $\Delta^2(k)=\rmd \sigma^2/
\rmd \ln k \propto k^3 P(k)$]. 
Our results indicate that $b(k)$ tends to a constant value on
scales where the density field is linear, with only a
weak tendency to change with scale for most reasonable bias
prescriptions: we shall denote this asymptotic large-scale bias
value by $b_\infty$.

Finally, we attempt to find a combination of a CDM-like mass model with a
reasonable normalisation and a
biasing prescription that can reproduce the observed power spectrum
of galaxy clustering, while reconciling observed cluster $M/L$ ratios
with the critical density assumed in Einstein -- de Sitter
models. Details of this exercise and its results are given in Section
5, and are discussed in Section 6, where we also present the
conclusions we draw from them.

\section{Eulerian  bias}

Our method is to choose some function to relate the galaxy and
mass densities
\beq
\rho_{\rm g}({\bf x})=f[\rho_{\rm m}({\bf x})],
\eeq
and investigate its biasing properties: we assume that these density
fields are defined on scales larger than that of individual galaxies.
This is straightforward where the matter density is
a continuous function, as is the case if there
is an analytical model for the density field.
The default model for the field at early times
is a random Gaussian process, but this must be
modified at late times when the field becomes
non-linear. The simplest model which introduces the
skewness necessary to avoid negative density is
a lognormal density field (Coles \& Jones 1991), for which the biased 
correlation function can be calculated analytically
in many cases.  These results are derived in the Appendix, and
yield simple expressions for the linear bias in the
limit of weak correlations: we shall
quote these below as
\beq
b_{\ss LN}\equiv \lim_{r \rightarrow \infty}\left[\frac {\xi_{\rm
g}(r)}{ \xi_{\rm m}(r)}\right]^{1/2}.
\eeq

More realistic non-linear distributions are provided by
$N$-body simulations, but here the density field
is discrete, and must be smoothed in order to
yield a continuous field. In practice, our
approach is to
weight the particles in the $N$-body simulation, 
depending on the local density of particles:
\beq
w=f[\rho_{\rm m}]/\rho_{\rm m},
\eeq
and to identify the
resultant weighted particle density with the local galaxy density.
Galaxies, of course, also form a discrete set of points, but we assume
galaxies to be a Poisson sample of some continuous field, whose
properties are to be estimated: we refer the reader to Coles (1993)
for a discussion of the Poisson clustering model and local bias.

From the infinity of possible relationships between
mass and galaxy density, we restrict ourselves to three parametric
forms which are nevertheless sufficiently flexible to allow
general conclusions to be drawn.  The first two model the effects of increasing
galaxy numbers in high density regions and suppressing 
them in voids, while the third is a functional form motivated by 
the numerical work of CO.
The functional forms we consider are detailed in the next three sections.

It is important to note that 
these models quantify the relationship between the mass and
galaxy distributions at redshift zero, rather than solely the process of
galaxy formation itself: as emphasised by CO and
others, the present-day distribution of galaxies reflects their
movement since birth as much as the locations of those births.
In particular, it may be difficult to predict the evolution of bias
(Matarrese et al. 1997).

\subsection{Power law bias}

The first model we consider is  power law bias:
\beq
\rho_{\rm g} \; \propto \;  \rho_{\rm m}^B,
\eeq
i.e. we apply a weight to each mass particle $\propto \rho_{\rm m}^{B-1}$.   
With $B>1$ equation (5)  represents the enhancement of galaxy numbers
in high density regions.
In the Appendix we show that the large-scale
bias parameter for power law biasing of lognormal mass models is
simply $b_{\rm LN}=B$.

A reasonable choice for $B$ would be the value required to decrease the 
mass-to-light ratio in clusters by a factor of about 5,
sufficient to reconcile $M/L$ measurements with an Einstein -- de Sitter 
universe.
If we take the centres of clusters to correspond to a mass overdensity of
$\sim 1000$, this suggests $B\simeq \ln(5000)/\ln(1000) = 1.23$, which
is perhaps a surprisingly small degree of non-linearity. This is only
a rough calculation, since it ignores the fact that the $\rho^B$
transformation also scales the mean luminosity density.
For the lognormal model in the Appendix, this correction
can be calculated, leading to the slightly higher value of $B \simeq 1.4$.
This argument is encouraging, since a small degree of bias
is exactly what is needed on large scales. The linear fractional
rms density contrast in $8\mpcoh$ spheres is about 0.9,
whereas the abundance of rich clusters suggests a figure of 0.5 -- 0.6
if $\Omega=1$ (White, Efstathiou \& Frenk 1993); a linear bias of 
about 1.5 is thus required.

\subsection{Censoring bias}

The next  prescription we study is censoring, in which we assume that galaxies
are not found in regions where the local mass density falls below a
certain value, $\tilde{\rho}_{\rm m}$: this represents both the
migration of galaxies from underdense regions and the possibility that
the galaxy formation process features some density threshold, below
which galaxies cannot have formed by the present. In the absence of other bias,
this is the `weighted bias'
model of Catelan et al. (1994), but it may readily be combined with
power law bias, to yield a model where galaxy numbers are enhanced in
high density regions and simultaneously suppressed in regions of low
density:
\beq
\rho_{\rm g} \propto 
		\left\{ \begin{array}{lll}
		\rho_{\rm m}^B & \mbox{ if $\rho_{\rm m} \geq 
				\tilde{\rho}_{\rm m}$} \\
		& \\          
		0 & \mbox{otherwise.}
		\end{array}
		\right.
\eeq
In the lognormal approximation the large-scale bias
for this model takes the form
\beq
b_{\rm LN} = B + \frac {\sqrt{2/\pi} \; {\rm e}^{-(\nu-B\sigma)^2/2}}
{\sigma \; \left[1 -{\rm erf} \left( \frac {\nu-B\sigma}{\sqrt{2}} \right) \,
\right].}
\eeq
where $\nu$ is the threshold for censoring in the Gaussian process from 
which the Eulerian density field is generated by a
lognormal transformation: $1+\delta_{\rm LN}=\exp(\delta_{\rm G}-\sigma^2/2)$,
where $\delta_{\rm LN}$ and $\delta_{\rm G}$ are the fractional
overdensities in the lognormal and Gaussian field, respectively;  
$\sigma^2$ is the variance of the latter; and erf  denotes the
error function. A more meaningful way to express the threshold is in
terms of the fraction, $f_{\rm M}$, of mass below the threshold, 
related to the Gaussian field by
\beq
f_{\rm M} =  0.5 \left[ 1 + {\rm erf}\left({\nu-\sigma\over\sqrt{2}}\right)\right].
\eeq
We shall generally use $f_{\rm M}$ to define density thresholds in
this paper, although, equivalently, one could use $f_{\rm V}$, the
fraction of the volume which is censored (i.e. given zero galaxy
density),  given by
\beq
f_{\rm V} = 0.5 \left[ 1 + {\rm erf}\left({\nu\over\sqrt{2}}\right)\right].
\eeq
In the case of $B=1$ equation (7) reduces to the
result  derived by Catelan et al. (1994) for their weighted
bias model, once account has been taken of the presence of an unwanted
factor of $1/\sqrt{2}$ in the definition of their quantity $\tilde{\nu}$.

In order to compare the large-scale bias level, $b_{\rm LN}$,
predicted for a particular bias scheme by the lognormal model with
that derived from a numerical simulation, we need to estimate $\sigma^2$.
Given that the density field is not exactly lognormal, the determination
of $\sigma^2$ will not be unique.   We estimate it by least-squares fitting 
to the differential form of (8),   using (9) to relate the parameter $\nu$
to the mass density: in what follows we show that the efficacy of the 
lognormal model may be judged independent of any problems in
estimating $\sigma^2$.

\subsection{Cen-Ostriker bias}

The final model we consider is inspired by the work of CO, 
who found a tight $\rho_{\rm g} - \rho_{\rm m}$
relation well fitted by the form 
\beq
\ln \left( \frac {\rho_{\rm g}}{\bar{\rho}_{\rm g}} \right) \; = \; C_0 +  
 C_1 \ln \left( \frac {\rho_{\rm m}}{\bar{\rho}_{\rm m}} \right)
+ C_2 \left[ \ln \left( \frac {\rho_{\rm m}} {\bar{\rho}_{\rm m}}
\right) \right]^2 .
\eeq
This is a generalisation of power law bias, with the
introduction of the second parameter allowing greater freedom for the
variation of the $\rho_{\rm m} - \rho_{\rm g}$ relationship: for
example, with $C_2 < 0$ one may mitigate the effects that a large power
law bias has in the densest environments.   Although Cen \& Ostriker found
specific values for $C_0$, $C_1$ and $C_2$,  we take them to be free 
parameters in this analysis.   The value
of $b_{\rm LN}$ for the CO models is given by
\beq
b_{\rm LN} \; = \; \frac {C_1-C_2 \sigma^2}{1-2 C_2 \sigma^2},
\eeq
as shown in the Appendix. This model has also been studied by 
Little \& Weinberg (1994), who investigated the void probability
function under CO bias.

\section{MODEL GALAXY POWER SPECTRA}

For each of the models considered, we compute a galaxy power spectrum
from the weighted particle distribution in a numerical simulation.   
The power spectrum is generally
still biased on the scale of the fundamental mode, so the galaxy power
spectrum is extended to larger scales by assuming a constant bias on
larger scales (see Section 3.2).   

\subsection{Small scales: numerical simulations}

Our numerical simulations were produced using the AP3M code of
Couchman (1991). All simulations were of an Einstein - de Sitter
universe, and used the CDM transfer function of BBKS:
\begin{eqnarray}
\lefteqn{T(k)  =  \frac {\ln (1 + 2.34q)}{2.34q} } \nonumber \\
& & \nonumber \\
& &  \times \left[ 1 + 3.89q + (16.1q)^2 + (5.46q)^3 + (6.71q)^4 
\right]^{-1/4} \!\!\!,  
\end{eqnarray}
where 
\beq
q \equiv k/\Gamma^*,
\eeq
wavenumbers, $k$, are in units of $\hompc$, the Hubble constant is
\mbox{$H_0 = 100 h \kmsmpc$}, and
$\Gamma^*$ is the shape parameter, which  for a pure CDM
universe with total and baryonic density parameters $\Omega_0$ and
$\Omega_{\rm B}$ respectively, is well fit (Sugiyama 1994)
by the form
\beq
\Gamma^*=\Omega_0 h \exp(- 
\Omega_{\rm B} - \Omega_{\rm B}/\Omega_0).
\eeq
(N.B. our shape parameter, $\Gamma^*$, is defined slightly
differently from the $\Gamma$ parameter introduced by Efstathiou,
Bond \& White 1992.) 

We consider $\Gamma^*$  to be a free parameter,
so that equations (12) and (13) describe a family of CDM-like mass power
spectra, \mbox{$\Delta^2_{\rm m}(k) \propto k^4 T^2(k)$}: we consider
only models with the Harrison-Zeldovich ($n=1$) primordial power
spectrum.

The form of these equations  means that a particular output from a CDM-like
simulation with a given value of $\Gamma^*$ may be identified with a
different epoch by reinterpreting the box side:
\beq
L_{\rm box}/\mpcoh \propto 1/\Gamma^*.
\eeq
Changing the assumed $\Gamma^*$ in this way also alters the
implied value of $\sigma_8$, so a single simulation output
gives results along a locus in $(\Gamma^*,\sigma_8)$ space.
Furthermore, different time-step outputs may be used to study 
different loci, so spanning the whole
family of CDM-like models at the present epoch. 
One cannot use
the earliest time-steps, which bear the traces of the initial
conditions, but this powerful technique can cover the interesting part of the
$(\Gamma^*,\sigma_8)$ space of mass models. The results in this paper come
principally from a
simulation of a $25/\Gamma^* \mpcoh$  box with 80$^3$ particles.

Biasing is implemented by weighting particles
according to their local density.  The local density must be calculated
quite carefully to avoid discreteness problems in low-density regions.
We divide the simulation box into $N^3$
cubic cells of side $\ell$, setting the density of the particles in a cell 
according to the number of particles in it.  
If, however, there are fewer than five 
particles in the cell, the cell side is repeatedly doubled until at least
five particles are found. 
  Essentially the assumption in this process is 
that there is no structure on scales smaller than the mean {\it local\/} 
inter-particle separation.
Inevitably there are some particles in very large void regions, and it
is computationally inefficient to extend this procedure of
doubling cell sizes until all of these are included in cells
containing at least five particles. As implemented here, this
procedure stops when $\sim 1$ per cent of the particles remain in
cells with less than five particles: the inaccuracy in the assignment
of local densities to these few particles in low density regions
has no bearing on the power
spectrum computed for the weighted particle distribution.

\bfi
\epsfig{file=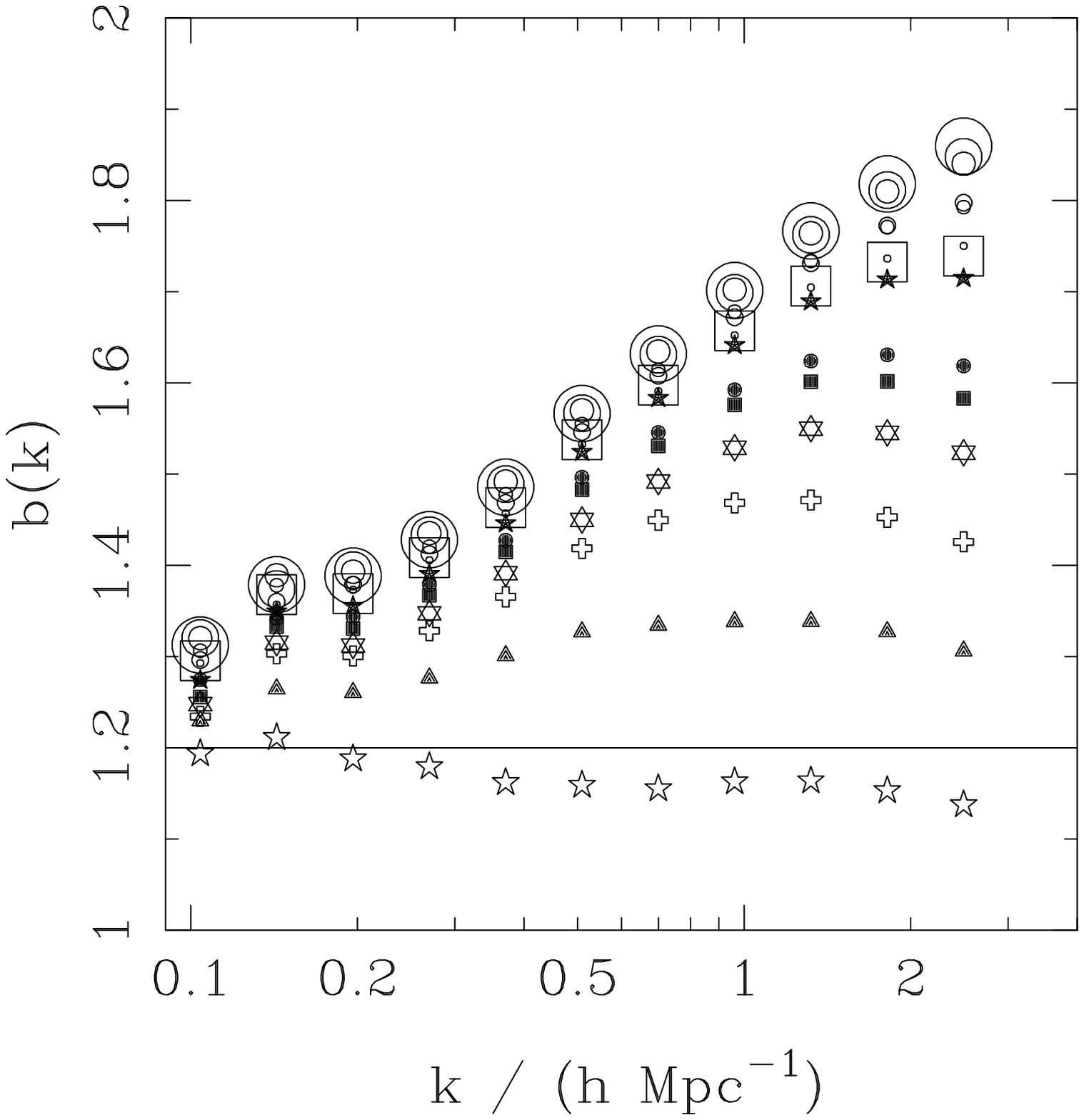,angle=0,width=8.5cm}
\caption{The scale-dependence of bias for power law bias
as a function of the resolution scale used to obtain the
local density: $[b(k)]^2$ is defined to be the ratio of `galaxy'
and mass power spectra.
For a mass model with $\Gamma^*=0.25, \; \sigma_8=0.64$,
we apply power law biasing with $B=1.2$. The various
symbols show the results of shrinking the cell used
to obtain the small-scale density contrast from $N$=10 (empty stars)
 to $N$=140 (largest circles). 
The horizontal line shows the linear bias value 
predicted by the lognormal model.
}
\efi

The degree of smoothing 
applied to the density field by this process clearly depends on the
value of $N$. In Fig. 1 we show how the level of bias
produced by a particular power law bias model from Section 4
varies with the value of
$N$: as one would expect, the small-scale bias increases with $N$, (as
less smoothing allows greater local densities, and, thus, higher
particle weights), until it saturates at a certain value of $N$, where
the cell size becomes smaller than the size of the most overdense
virialized regions, and further reduction cannot increase the smoothed densities
produced. This saturation sets in at $N \simeq100$, and we choose 
for the evaluation of our local densities a value of $N=120$,
corresponding to \mbox{$\ell=0.83 (0.25/\Gamma^*) \mpcoh$}. 
For the mass models we are most interested in,
with $\Gamma^* \simeq 0.25$, this is roughly equivalent to smoothing
with a spherical top hat of filter radius \mbox{$R_{\rm TH} \simeq 0.5
\mpcoh$}, so that the scale on which we define our local density
is similar to that on which cluster $M/L$ ratios are evaluated, and
on which we assume  physical bias mechanisms operate. Note that, in
Fig. 1 and subsequent figures, a numerical artefact means that the 
power in the longest-wavelength bin is slightly incorrect: the
magnitude of this effect is small enough to have no bearing on our
results.

\subsection{Large scales: quasi-linear power spectrum extension}

The range of scales which we need to consider is too large for present-day
$N$-body simulations, so we use the method of 
PD94 to map linear power to
non-linear power on large scales.  The heart of this method,
based on that of Hamilton et al. (1991) for evolving the
integrated correlation
function $\bar{\xi}(r)$, is the pair of postulates that: (a) gravitational
collapse can be considered as a translation of scales, so that
wavenumbers $k_{\ss L}$ and $k_{\ss NL}$ in the linear and
non-linear power spectrum, respectively, are related by
\beq
k_{\ss L} \; = \; \left[1 + \Delta^2_{\ss NL}(k_{\ss NL}) 
\right]^{-1/3} k_{\ss NL},
\eeq
where $\Delta^2_{\ss NL}(k_{\ss NL})$ is the non-linear power
spectrum; and (b) that there should be a universal mapping
\mbox{$\Delta^2_{\ss NL}(k_{\ss NL}) \; = \; f_{\ss NL} \left[
\Delta^2_{\ss L}(k_{\ss L})\right]$}
between the linear and non-linear power spectra. PD94 used
knowledge of the asymptotic behaviour of $f_{\ss NL}$,
together with numerical simulation data, to determine its
full form. A more recent paper (Peacock \& Dodds 1996) 
takes account of a weak dependence
of $f_{\ss NL}$ on the linear power spectrum, but the difference
between the two prescriptions is small on the quasi-linear scales in which
we are interested.

To perform this extension, we take the bias parameter $b(k)$ to 
be constant on large scales.  This is motivated by the numerical
results below, which show that $b(k)$ indeed tends to a constant at small $k$.
We estimate the constant from the smallest three wavenumber
bins used in the determination of the power spectrum of the weighted
particle distribution in the numerical
simulation, and tests on larger boxes suggest that this procedure for 
estimating the asymptotic value of $b(k)$ is 
accurate to $\sim 0.1$ or better for the bias prescriptions in which
we are
most interested. An alternative method for setting the large-scale 
bias would be to use the analytic lognormal model,  but, as shown in 
Fig. 2 and subsequent figures below, this is not usually sufficiently accurate.

\bfi
\epsfig{file=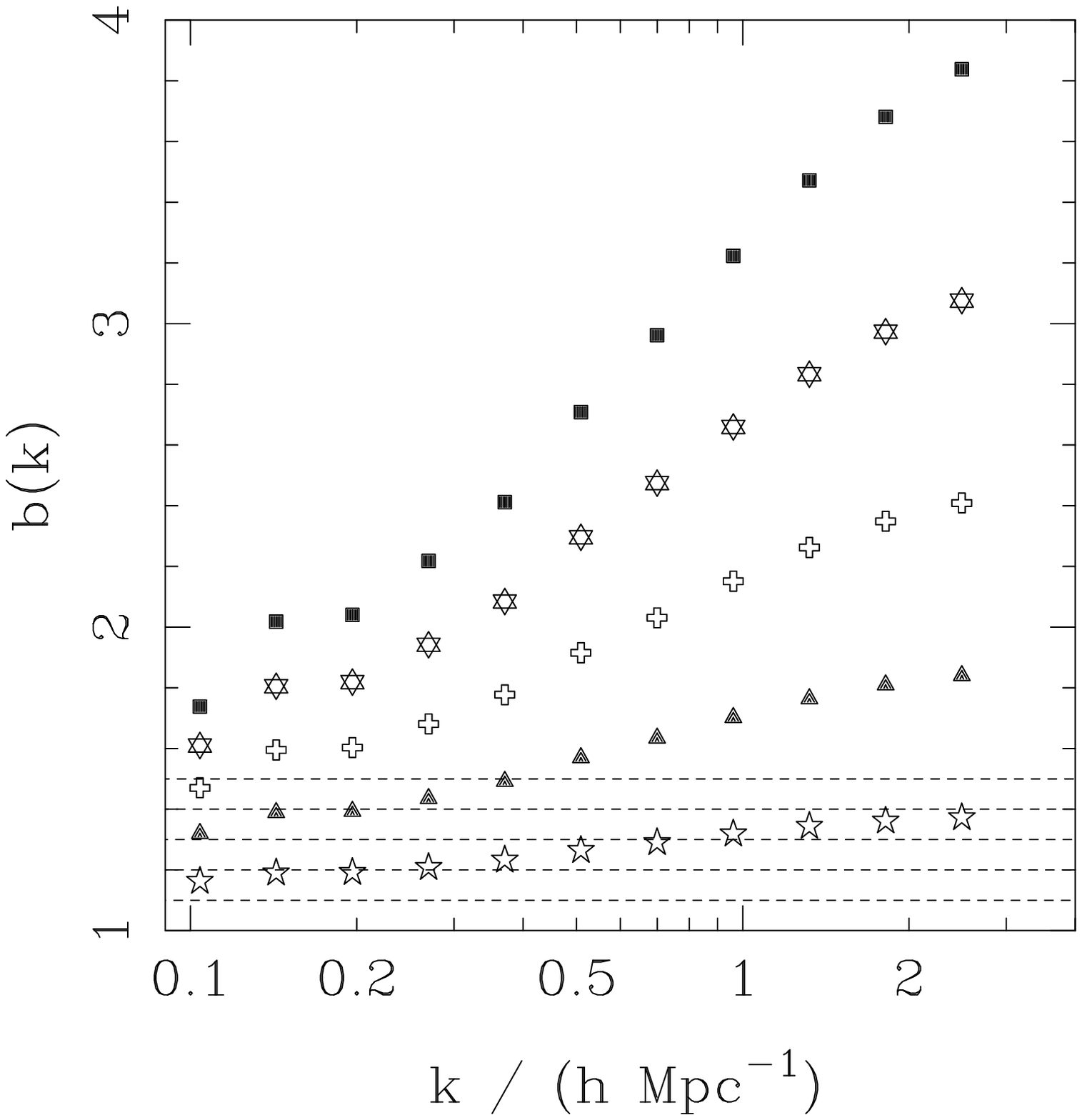,angle=0,width=8.5cm}
\caption{The scale-dependence of bias for power law bias
as a function of the power law index:
$B=1.1, 1.2, 1.3, 1.4, 1.5$ in order of increasing bias.
A CDM model with $(\Gamma^*=0.25,\sigma_8=0.64)$ is assumed.
The horizontal lines show the linear bias values
predicted by the lognormal model.
}
\efi

\section{POWER SPECTRA WITH EULERIAN BIAS}

\subsection{Power law bias}

In Fig. 2 we show the $b(k)$ curves for a range of $B$ values and a
single mass model $(\Gamma^*=0.25,\sigma_8=0.64)$.
As one would expect, the $b(k)$ curves steepen as $B$ gets
larger: giving more weight to over-dense regions increases the overall
level of bias, as these regions are more strongly clustered than the
mass distribution as a whole, but this is felt most strongly at high
$k$, which probes clustering power on  separations within the scale of
these regions themselves.
The dashed lines show the corresponding values of $b_{\rm LN}$,
which appear to yield a slight underestimate of the asymptotic
level of bias.

\bfi
\epsfig{file=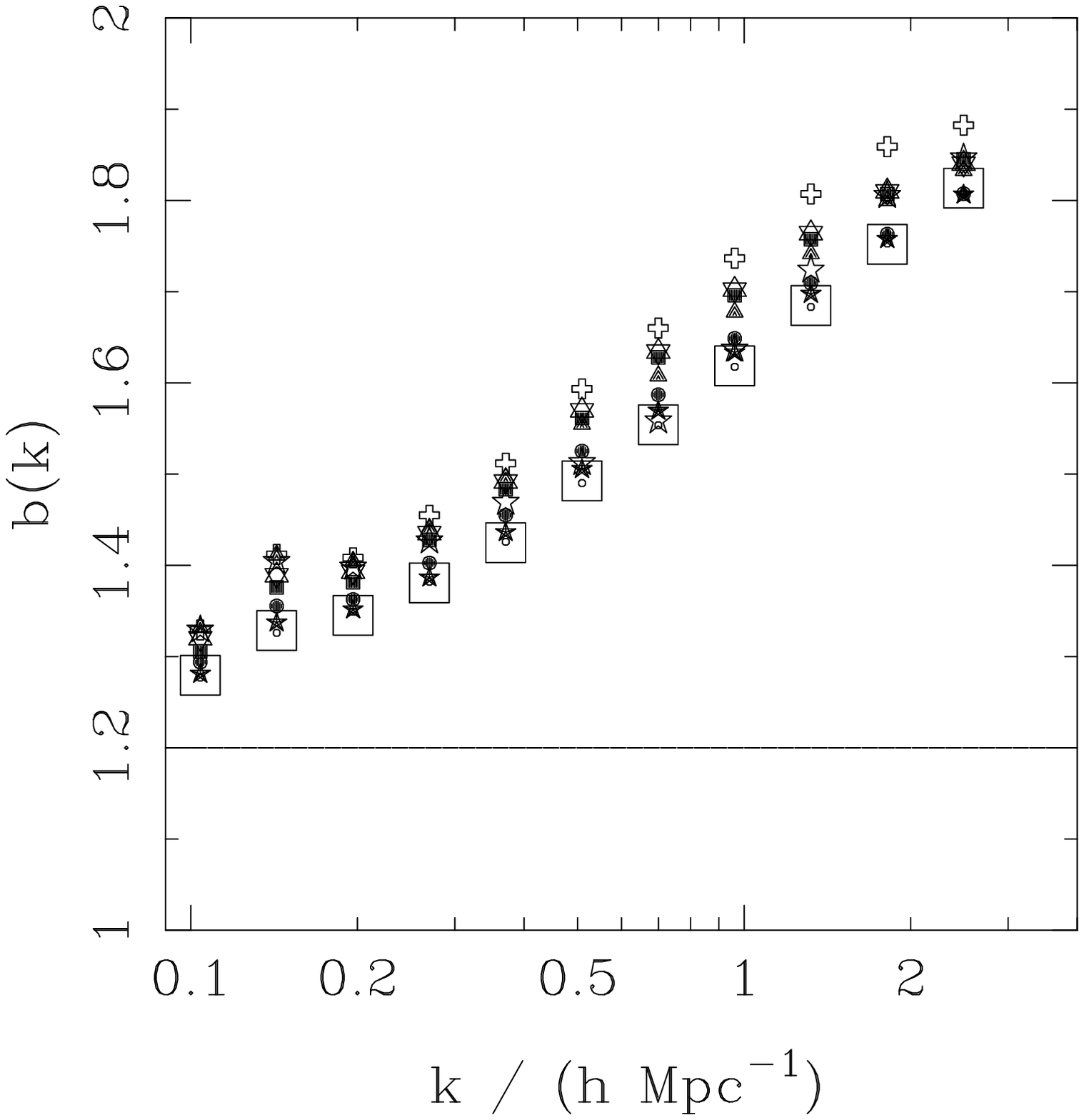,angle=0,width=8.5cm}
\caption{The scale-dependence of bias for power law bias
with $B=$1.2, for CDM models with $\Gamma^*=0.25$
and $\sigma_8$ varying from 0.37  to 1.0. The $\sigma_8$ values
plotted are as follows: 0.37 (empty stars), 0.47 (triangles), 0.56
(crosses), 0.64 (stars of David), 0.72 (filled squares), 0.80 (filled
circles), 0.87 (filled stars), 0.93 (empty squares), and 1.0 (empty circles).
Note that the trend of bias at given $k$ is not
monotonic with $\sigma_8$, and that the maximum bias is attained
at $\sigma_8=0.56$: this is discussed further in the main text.
The horizontal line shows the linear bias value 
predicted by the lognormal model.}

\efi

\bfi
\epsfig{file=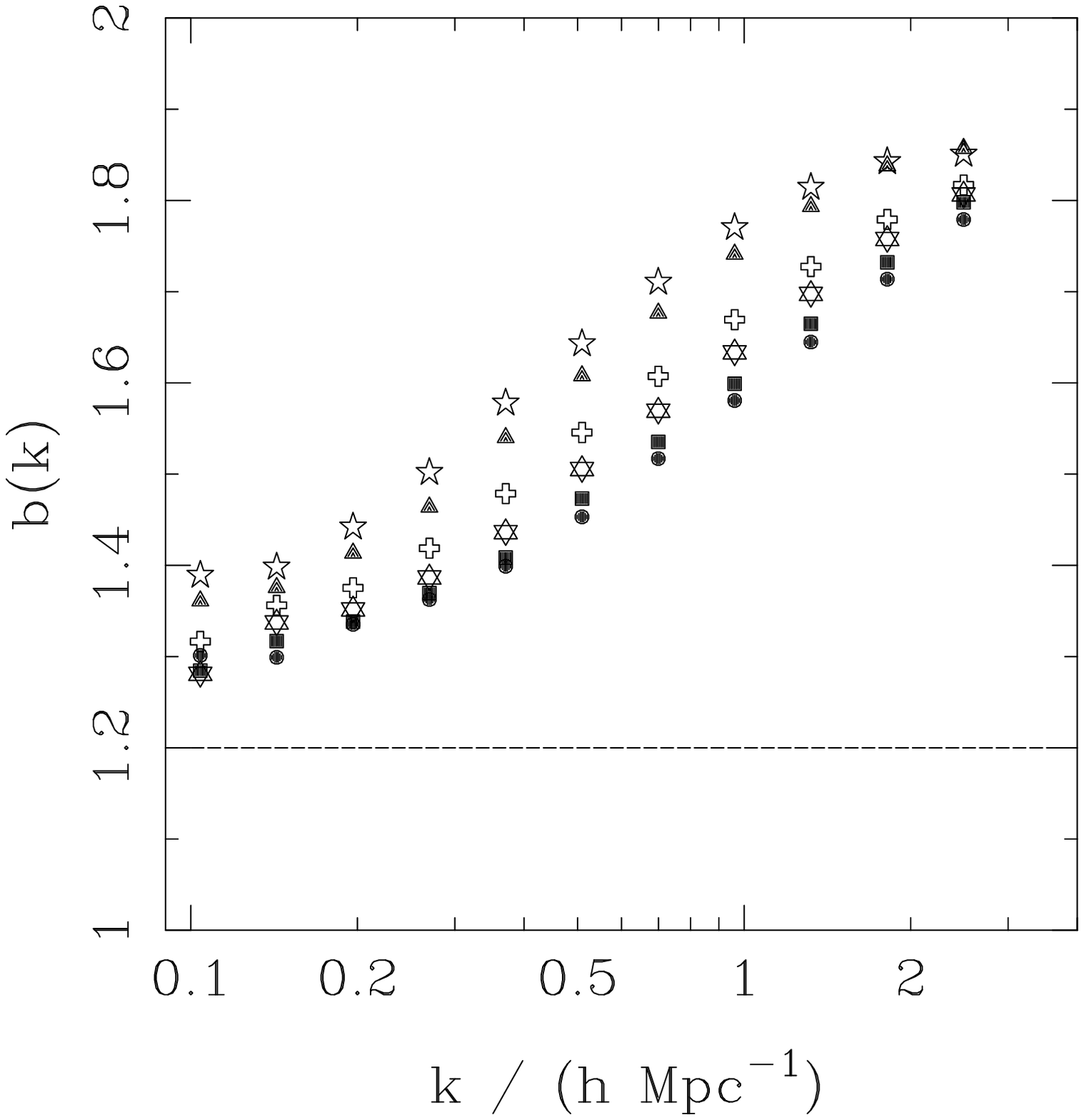,angle=0,width=8.5cm}
\caption{The scale-dependence of bias for power law bias
with $B=$1.2, for CDM models with $\sigma_8\simeq0.85$
and $\Gamma^*$ varying from 0.175 (highest) to 0.3 (lowest).
The horizontal line shows the linear bias value 
predicted by the lognormal model.}
\efi

Figs. 3 \& 4 show, respectively, the effect on $b(k)$ of varying the amplitude
and shape of the linear mass power spectrum.
From Fig. 3 we see that $b(k)$ is remarkably
insensitive to amplitude: the small-scale bias varies by only about
15 per cent between the earliest and latest epochs, although the power
in those wavenumber bins increases by more than a factor of ten in
that time.  
Detailed inspection of Fig. 3 shows that
$b(k)$ at given $k$ initially increases with increasing 
$\sigma_8$ to $\sigma_8 \simeq 0.6$, after 
which the amplitude of the $b(k)$ curve begins to fall as $\sigma_8$
increases further. One would expect $b(k)$ to increase with
$\sigma_8$, since, as $\sigma_8$ gets larger, more mass becomes
concentrated in the densest environments, which increases the level of
bias on all scales, but preferentially on small scales, for the
reasons given above to explain Fig. 2. To understand the reduction in
$b(k)$ at large $\sigma_8$, consider what would happen if the
simulation had been left running for a long time, until $\sigma_8 \gg
1$. In the limit of a very highly evolved simulation, essentially all the
mass would reside in a number of very massive clumps. Applying bias to
such a mass distribution would have very little effect, so we expect 
\mbox{$b(k) \rightarrow 1, \;\; {\rm as} \;\; \sigma_8 \rightarrow
\infty$}, hence a reduction in $b(k)$ for larger $\sigma_8$.
What this tells us is that the latest epochs in the simulations
are starting to become affected by the absence of 
very large-scale power beyond the box scale.

In Fig. 4 we see a slightly larger spread in $b(k)$, with models
with lower $\Gamma^*$ having somewhat larger bias at all
wavenumbers. Given the form of the transfer function of CDM-like
models (equations 12 and 13), we see that the effect of varying
$\Gamma^*$ at constant $\sigma_8$ can be thought of as shifting the
power spectrum in wavenumber, to a model with a different degree of
evolution, which is lower for the lower $\Gamma^*$ values. Thus, one
can imagine producing Fig. 4 by shifting the $b(k)$ curves in Fig. 3
to larger $k$: the value of
$\sigma_8$ used for the models plotted in Fig. 4 is larger than the
critical value of $\sigma_8 \simeq 0.6$, so the Fig. 3 curves one would
shift would all be flattening ones in the higher $\sigma_8$
regime. In this way, we see that the variation in $b(k)$ with
$\Gamma^*$ shown in Fig.4 reflects how concentrated into large,
discrete clumps is the mass in a CDM-like model at constant $\sigma_8$
as a function of $\Gamma^*$, which determines how large an effect is
produced by biasing that mass distribution.

Overall, the main features of these plots are that 
the bias is a monotonic function of scale and that it
varies relatively slowly as a function of scale,
changing from e.g. 1.5 to 3.0 over a range of
scales where the observed $\Delta^2(k)$ alters by
a factor of more than 300.
The slow variation of $b(k)$ is in marked
contrast to the lognormal prediction:
$\smash{1+\xi_{\rm g}=\left(1+\xi_{\rm m}\right)^{B^2}}$, where
$\xi_{\rm m}$ and $\xi_{\rm g}$ are, respectively, the mass and galaxy
correlation functions. For $B=1.5$, this implies an effective bias
of 75 at $\xi=1000$, which is a far more rapid variation
than is seen in the $N$-body results. Note also that
the lognormal model predictions indicated by the horizontal lines
in Figs. 2-4 always appear to lie below the asymptotes towards which the
numerical $b(k)$ curves are tending.

\subsection{Cen-Ostriker bias}

The $b(k)$ curves resulting from biasing the mass model of Fig. 1 by a
series of CO models are shown in Fig. 5:
the value of $C_1$ is held
constant and $C_2$ is varied, to show the sensitivity of the model
galaxy power spectrum to the weighting of the very densest regions.
Note that, for $C_2<0$, the suppression of the weight given to
the densest regions reduces the overall level of bias, so that the 
large-scale
asymptote, $b_\infty$, falls below $b_\infty=C_1$, and that there is
a change in the shape of the $b(k)$ curve, making $b(k)$
increase with scale. 
The lognormal predictions are shown on Fig. 5, and
capture the main effect of varying $C_2$: nevertheless, as with pure
power law bias, the model does not predict the asymptotic level of
bias perfectly.

Coles (1993) showed that local biasing of a density field which
is Gaussian, or a local transformation thereof, will produce a
real-space bias parameter, $b(r)$, (defined to be the square root of
the ratio of the galaxy and mass correlation functions)
which decreases monotonically with scale. For CDM-like spectra, the
same analysis predicts a monotonic decrease in $b(k)$ with increasing
scale, with the
exception of the very largest scales \mbox{($k \la 0.01 \; h$ Mpc$^{-1}$)}, 
where the tangent spectral index is $n_{\rm eff}>0$ and, consequently,
$b(k)$ can increase unboundedly. The results of Fig. 5 show how
the Coles (1993) inequalities  breaks down once the mass distribution becomes
sufficiently non-linear; like the failure of the lognormal model to
predict correctly the large-scale asymptotic bias level, this indicates that
the evolved mass distribution is not well described by a lognormal.

\bfi
\epsfig{file=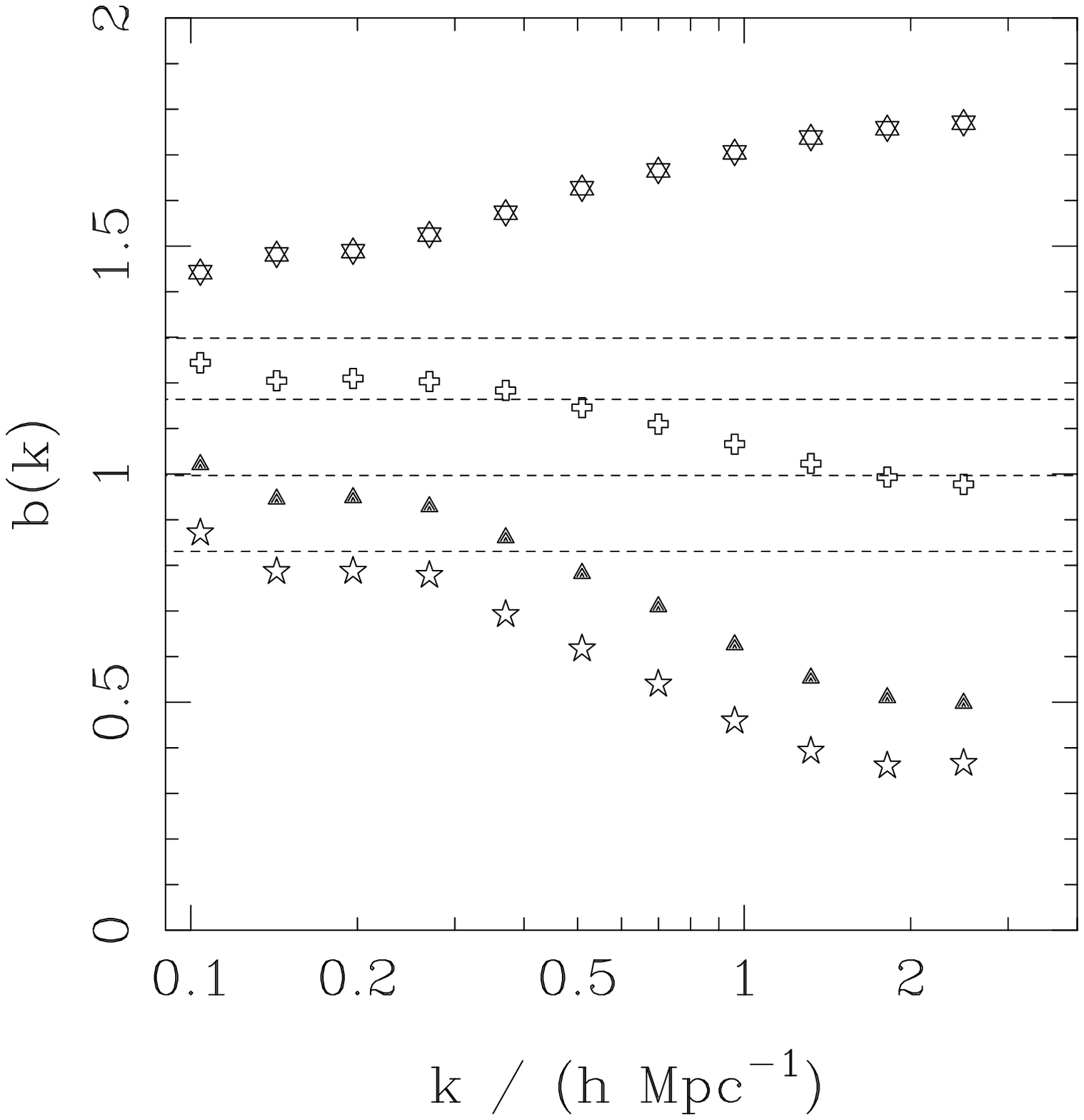,angle=0,width=8.5cm}
\caption{The scale dependence of bias for CO bias
with $C_1=$1.5 and $C_2$ varying from -0.4 (lowest) to -0.05
(highest), for a mass model with $\Gamma^*=0.25$ and $\sigma_8=0.64$.
The horizontal line shows the linear bias value 
predicted by the lognormal model.
}
\efi

\subsection{Censoring}

The result of censoring the same mass model is shown in Fig. 6, which
indicates that 
the level of bias increases as more of the void regions are removed. 
However,  it is only when \mbox{$f_{\rm M} \ga 0.6$} 
that the curvature in $b(k)$ begins to increase appreciably,
showing again how the small-scale power is determined by the densest
regions of space. The same can be said for Fig. 7, which shows the
same effect in the censoring of a $B=1.2$ power law bias model at
various thresholds. The dashed lines are the predictions of the
lognormal model, which once again produces a fair, but not perfect,
estimate of the level of large-scale bias.

Both in the lognormal approximation, and in the numerical datasets,
the effects of censoring seem less extreme than might have been
expected on the basis of the following naive argument. 
If we make up the mass distribution by the addition
of an unclustered background to a clustered field, then censoring
might be expected to correspond roughly to the removal of
the unclustered part. If we write
\beq
\bar\rho (1+\delta) = \bar\rho_c (1+\delta_c) + \rho_u,
\eeq
then we have
\beq
\delta_c=[1+\rho_u/\bar\rho_c]\, \delta,
\eeq
so that the effective real-space bias is just $b=(1-f_{\rm M})^{-1}$, where 
$f_{\rm M}=\rho_u/(\bar\rho_c+\rho_u)$ is the fraction of the
mass which has been censored. To solve the $M/L$ problem
by censoring alone, we would need $f_{\rm M}\simeq 0.8$, implying
$b=5$ by this argument. However, the lognormal model
yields only \mbox{$b=1.56$} for $f_{\rm M}=0.8$ (assuming
$\sigma=2.5$, but only
weakly dependent on $\sigma$): so, an amount of censoring
large enough to have an important effect on $M/L$ values
thus produces the sort of bias factor needed in reality 
to give a sensible $\sigma_8$ if $\Omega=1$.

\bfi
\epsfig{file=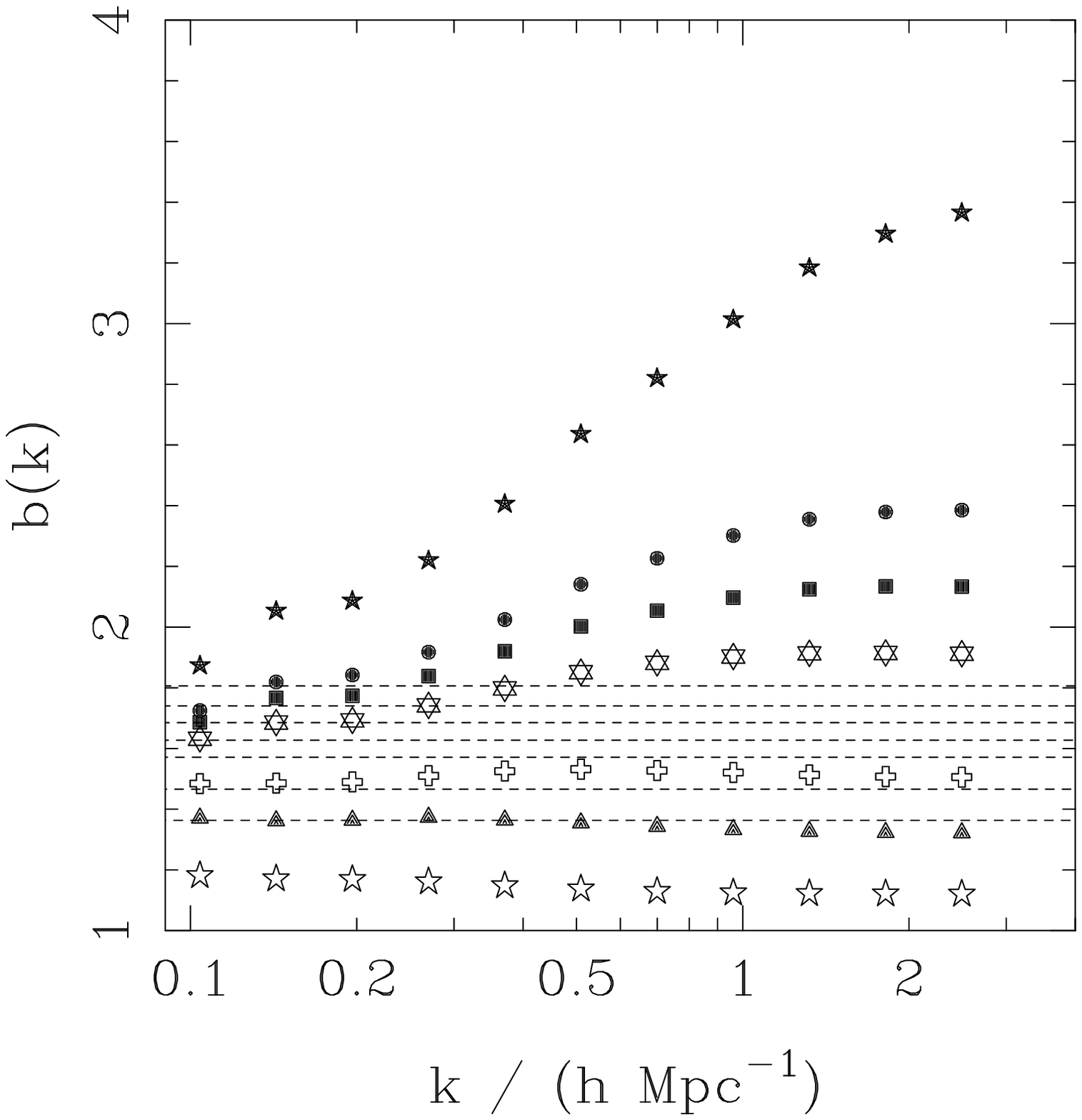,angle=0,width=8.5cm}
\caption{The scale-dependence of bias for censoring bias
with $f_{\rm M}$ varying from 0.35 (lowest) to 0.65 (highest), for a
mass model with $\Gamma^*=0.25$ and $\sigma_8=0.64$.
The horizontal lines show the linear bias values
predicted by the lognormal model.
}
\efi

\bfi
\epsfig{file=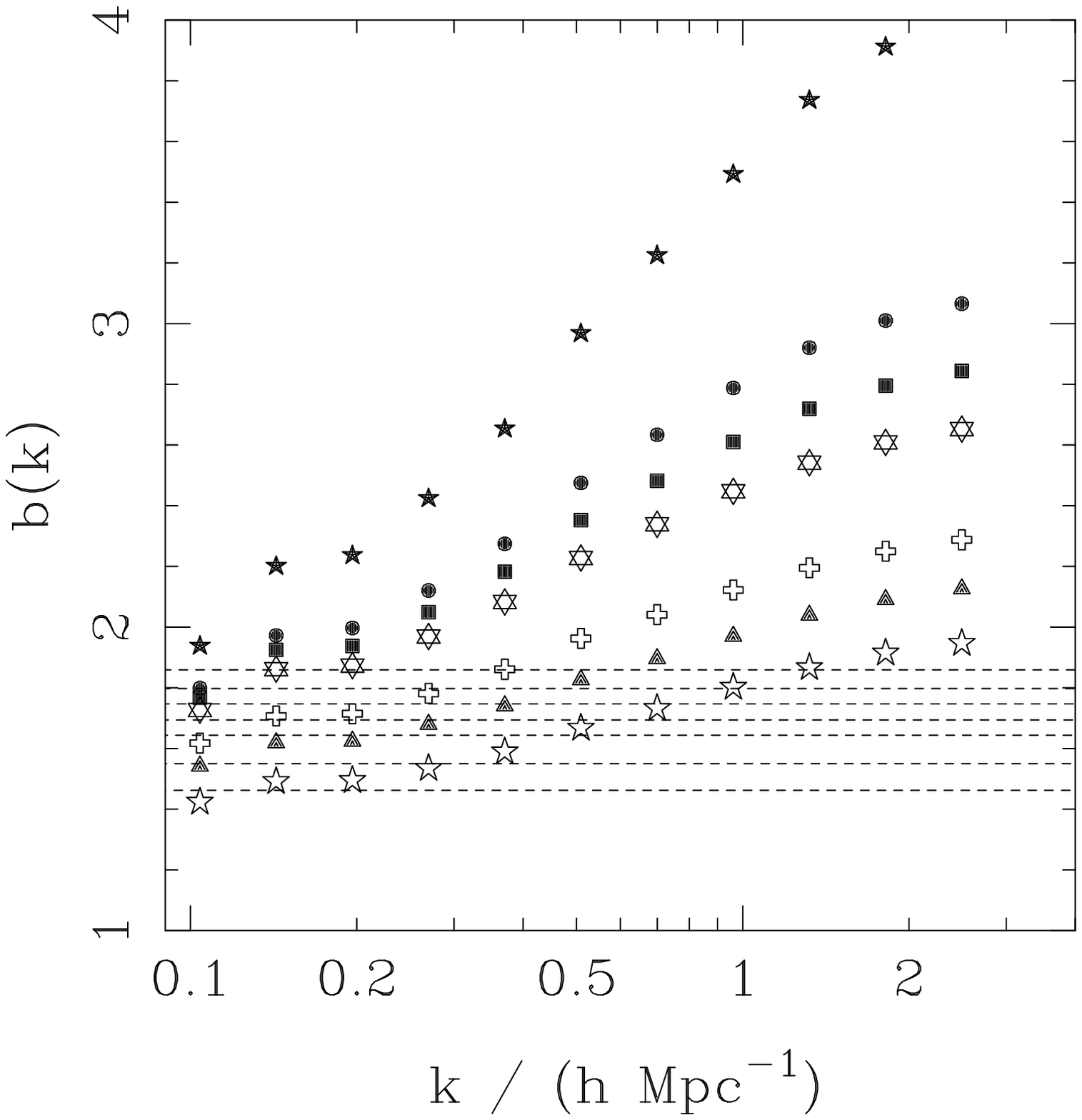,angle=0,width=8.5cm}
\caption{The scale dependence of bias for a
combination of power law bias with $B=1.2$ and censoring
with $f_{\rm M}$ varying from 0.35 (lowest) to 0.65 (highest), for a
mass model with $\Gamma^*=0.25$ and $\sigma_8=0.64$.
The horizontal lines show the linear bias values
predicted by the lognormal model.
}
\efi

\bfi
\epsfig{file=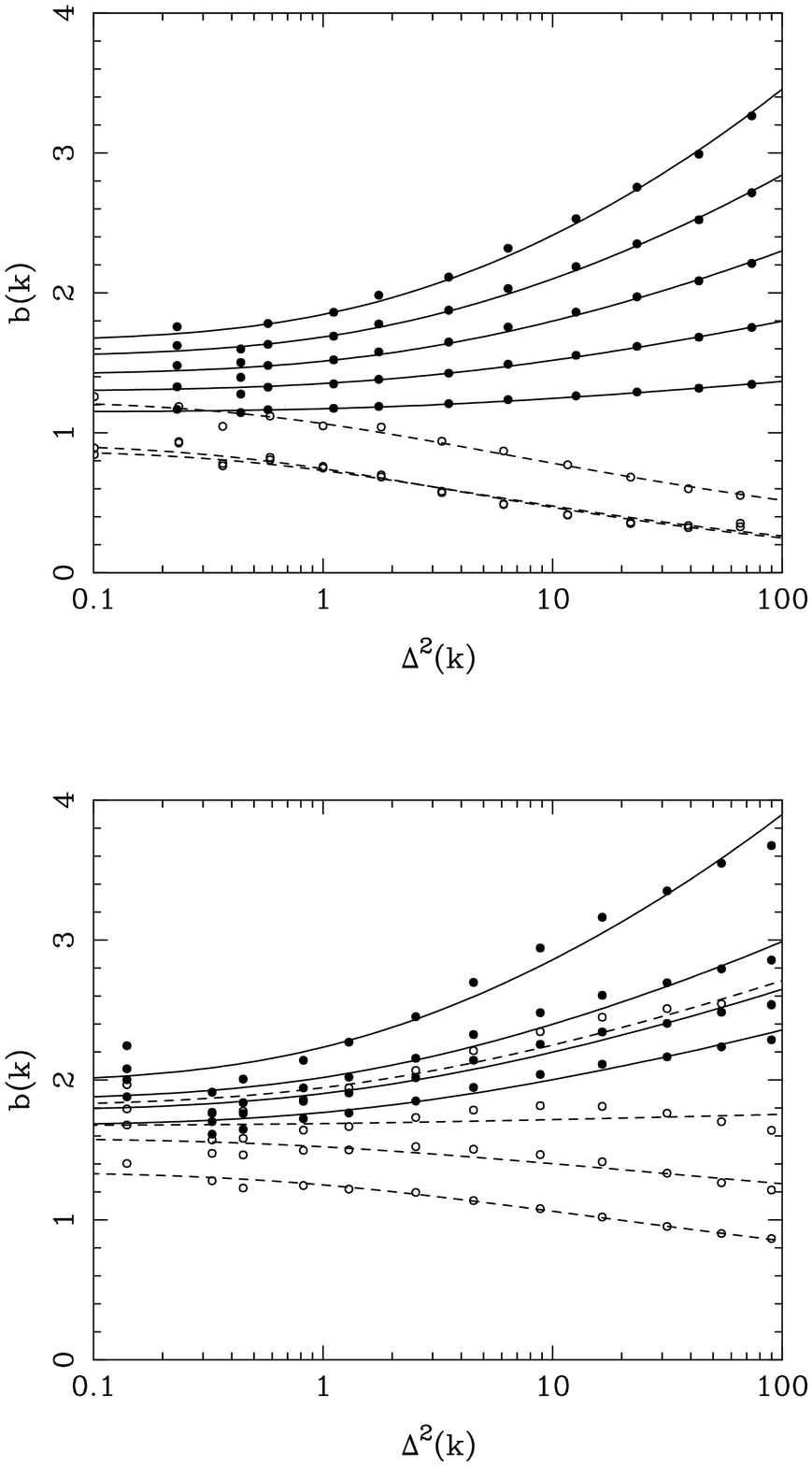,angle=0,width=8.5cm}
\caption{Numerical data for $b(k)$ and model fits.
This figure considers the case $\Gamma^*=0.25, \sigma_8=1$ only.
The top panel shows pure power law bias with $B=1.1, 1.2, 1.3, 1.4, 1.5$
(solid points and full lines) and CO bias with $C_1=1.2$ and
$C_2=-0.1, -0.4, -0.7$ (open points and dashed lines). The bottom
panel shows a combination of censoring ($f_{\rm M}=0.3, 0.4, 0.5, 0.6$)
with power law bias: $B=0.8$ (solid points and full lines)
and $B=1.2$ (open points and dashed lines).
}
\efi

\subsection{Scale dependence of bias}

One interesting general feature of the bias seen
in the numerical results is the relatively weak scale
dependence. Consider the prediction in
case of power law bias of a lognormal field, as given in the Appendix:
\beq
1+\xi_g = (1+ \xi_m)^{B^2}.
\eeq
This implies a very rapid increase of bias, once the
field reaches scales where the fluctuations are non-linear.

Even for the case of power law bias, such a rapid dependence
on scale is not seen in our $N$-body data, as illustrated
for a variety of models below in Fig. 8.
This plot shows a fitting formula for the scale-dependence
of bias in $k$ space, which is inspired by the lognormal
form, but which introduces an extra parameter in order to allow
pure scale-independent bias as one limit:
\beq
1+\Delta_{\rm g}^2 = (1+b_1 \Delta_{\rm m}^2)^{b_2}.
\eeq
As may be seen from the figure, this form is capable of
giving an excellent description of the data.
For  pure power law bias, the results scale with $B$
almost exactly as
\beq
(b_1, b_2) = (B^{1.8},B^{0.7}),
\eeq
so the non-linear component of the bias is much smaller
than the scale-independent component: this is a general
feature of all the models we have investigated.

Notice also that the power law models have a large-scale
linear bias of $b_\infty=[b_1 b_2]^{1/2}\simeq B^{1.3}$, which is larger
than the simple relation $b_\infty=B$ expected from the lognormal
model. The lognormal model therefore fails quite badly as a
model for the non-linear density field produced in 
strong gravitational non-linear evolution, which
should come as no surprise: the models studied in the
Appendix are invaluable as heuristic aids, but should not be
applied quantitatively to the real Universe.

\section{COMPARISON OF CDM-LIKE MODELS WITH OBSERVED GALAXY CLUSTERING}

We now attempt to use some of the above results to
investigate which models are capable of accounting
for the observed statistics of galaxy clustering.
This can be approached at a number of levels, of
which the simplest is the study of the fluctuation
spectrum. We shall mainly restrict ourselves to this
question here, but it may be useful to begin by
illustrating in more detail the effect of our
bias models on the simulated galaxy density field.

\bfi
\epsfig{file=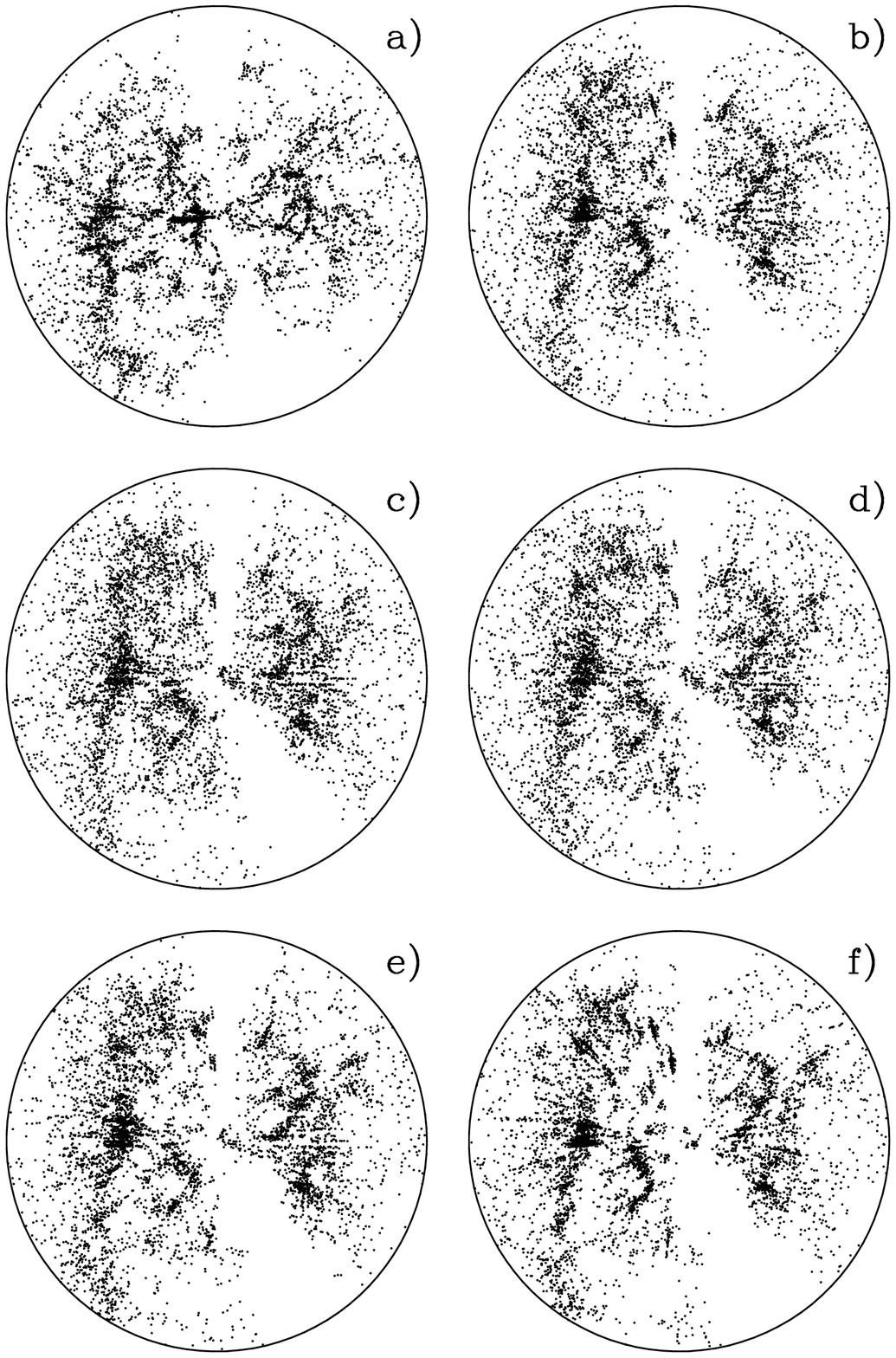,angle=0,width=8.5cm}
\caption{A plot of the CfA slice [$30^\circ < \delta < 60^\circ$;
$V<12000 \kms$] (panel a), with a variety of synthetic $N$-body
simulations having the same selection function and numbers
of `galaxies'. The simulations used $n=-1.5$ initial
conditions, with $\Omega=0.3$ in panel (b) and
$\Omega=1$ in panel (c). The normalization was set at $\sigma_8=1.0$
in (b) and 0.6 in (c).
Panels (d) and (e) bias the $\Omega=1$ data with respectively
$B=1.23$ and $B=1.23, f_{\rm M}=0.1$. Panel (f) biases the open
simulation with $f_{\rm M}=0.1$. It appears that censoring is
needed in both cases to obtain sufficiently empty voids.
}
\efi

Fig. 9 compares the true local galaxy distribution with various models.
In panel (a) we show the local ($V<1200 \kms$) distribution of galaxies
from the CfA slice (Huchra et al. 1992, with more recent updates), while
simulations (b) and (c) show unbiased $\Omega=0.3$ and $\Omega=1$
universes, respectively, for a mass model whose linear power spectrum
has a slope ($n=-1.5$) roughly equaling that of CDM on galactic scales 
and a normalization
set by the cluster abundances. The clearest feature
of these examples is an insufficient contrast between high density
regions and voids. Panel (d) applies power law bias with $B=1.23$ to the $\Omega=1$ case, but this improves the
situation only slightly. Panel (e) adds censoring with
$f_{\rm M}=0.1$, and this `opens out' the voids to roughly the
correct extent. Finally, panel (f) applies the same
censoring to the low density model, again with
reasonable results at the visual level.
We thus believe that our bias models
do capture the main essentials of the modifications needed
in order to make dark matter distributions resemble the
real distribution of galaxies.
For the remainder of the present paper, we shall be content
to restrict ourselves to comparison of data and model
at the two-point level.

\subsection{Observational data}

The observational data we use for comparison with the models are: 
\begin{description}
\item{(i) Large-scale normalisation:  the four-year {\em COBE} data yield
 \mbox{$Q_{\rm rms-PS}=
18.4 \mu K$}  (e.g. Tegmark 1997). We allow a conservative 10 per 
cent uncertainty in that figure.}

\item{(ii) Intermediate scale normalisation: $\sigma_8 = 0.57 \pm 0.05$
from the abundance of rich clusters (White et al. 1993).} We consider models
with \mbox{$0.5 < \sigma_8 < 0.7$}.

\item{(iii) Galaxy clustering:  real-space APM
galaxy power spectrum (Baugh \& Efstathiou 1993) over the
wavenumber range \mbox{$-1.5 \leq \log_{10}(k/ \hompc)
\leq 0.5$}.}

\end{description}

The galaxy clustering range chosen spans the apparent point of
inflection in the
APM power spectrum, including data on the linear scales studied by
PD94, as well as the non-linear regime omitted from their analysis.  We have
not used the largest scales because of possible uncertainties in surface
uniformity of the survey, and have multiplied the power spectrum of 
Baugh \& Efstathiou (1993) by
1.25 to account for the non-zero median redshift of
the APM survey: see Peacock (1997) for details of that correction.

For models which satisfy the above constraints, we investigate whether
they can give sufficiently large mass-to-light ratios in clusters. 
In rough terms, we require \mbox{$\rho_g/\bar\rho_g \simeq 5000$} in regions
where  \mbox{$\rho_m/\bar\rho_m \simeq 1000$}.
One can argue in detail about these numbers, but they should be
approximately correct for cluster masses and sizes: in fact, as we shall
see, no model gets close to the required galaxy density.

We note in passing that we find that no $\Gamma^*$ model can be linearly
biased [i.e. $\Delta^2_g = b^2(k) \Delta_m^2$, with $b(k)$ constant] in such
a way as to satisfy these three constraints.
The problem lies with the shapes of the evolved mass power
spectra of CDM-like models, which do not match that of the APM power
spectrum, implying that to bias successfully a $\Gamma^*$ model
will require a non-linear $b(k)$, and thus motivating our study of non-linear
bias prescriptions.  In any case, it is not clear how a linear bias may
be achieved from a local transformation of a highly non-linear density field,
although numerical studies involving detailed prescriptions for
galaxy formation have shown a nearly linear bias (Kauffmann, Nusser \&
Steinmetz 1997).

\bfi
\epsfig{file=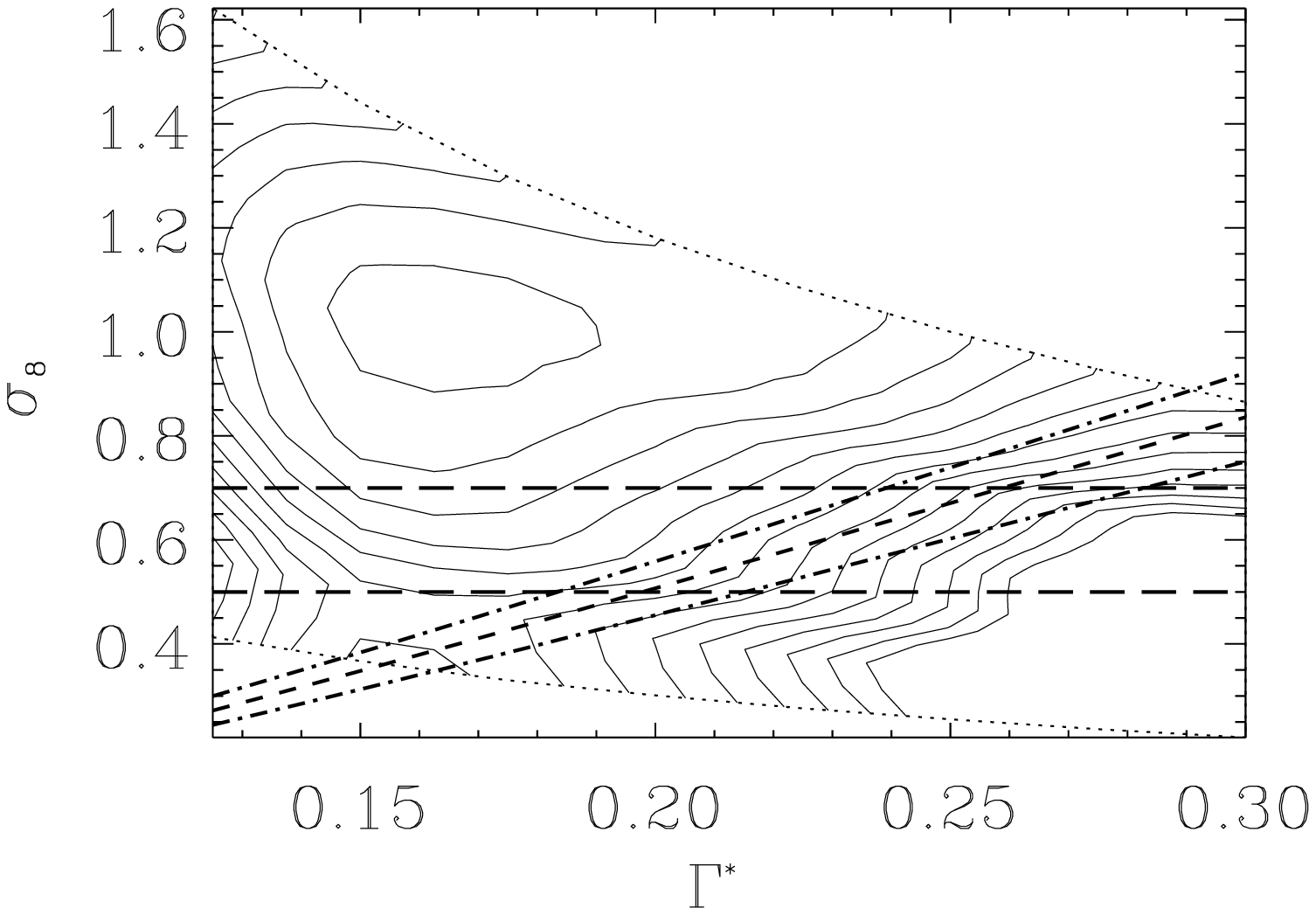,angle=0,width=8.5cm}
\caption{Contours of $\chi^2$ between power law biased CDM power spectra
and the APM data: the contours are at $\Delta \chi^2=15$, with
$\chi^2=30$ being the lowest contour shown, near the minimum at $\Gamma^* 
\simeq 0.17,$ $\sigma_8 \simeq 1.0$.
The horizontal lines show the allowed range of $\sigma_8$ from the
cluster abundance constraint
The diagonal lines show the constraint required to
satisfy the {\it COBE\/} measurement for the case of no tilt.
}
\efi

\subsection{Results}

\subsubsection{Power law bias}

In Fig. 10 we show contours of $\chi^2$ for fits of power law
biased CDM-like models to the APM data.
There is a clear preferred model, with
$\Gamma^*= 0.175, \sigma_8=1.03, B=0.86$ and
$\chi^2=14.5$, for 13 degrees of freedom: the power spectra are
compared at 14 wavenumbers, so there are 13 degrees of freedom in
the power law bias fits and 12 in the $(B,f_{\rm M})$ and $(C_1,C_2)$ fits.
This model is however firmly excluded both by the cluster abundance 
constraint (horizontal lines) and {\em COBE} (angled 
lines) if the primordial spectral index is $n=1$.  This latter 
constraint is removed if the spectrum is tilted, but even so the
cluster abundance difficulty remains.
If we insist on a normalization of $\sigma_8\simeq0.6$, to fit the
cluster abundance normalisation, the
preferred value of $\Gamma^*$ is unchanged at 0.175, 
$B$ increases to 1.1 and $\chi^2$ increases sharply to 43.3.
These models are compared to the APM data in Fig. 11.
Neither alternative fully reproduces the observed
data, especially those below the inflection around $k\simeq 0.1 \hompc$,
but they exploit the errors in order to reduce the
discrepancy to the point where the overall $\chi^2$
is moderate.

\bfi
\epsfig{file=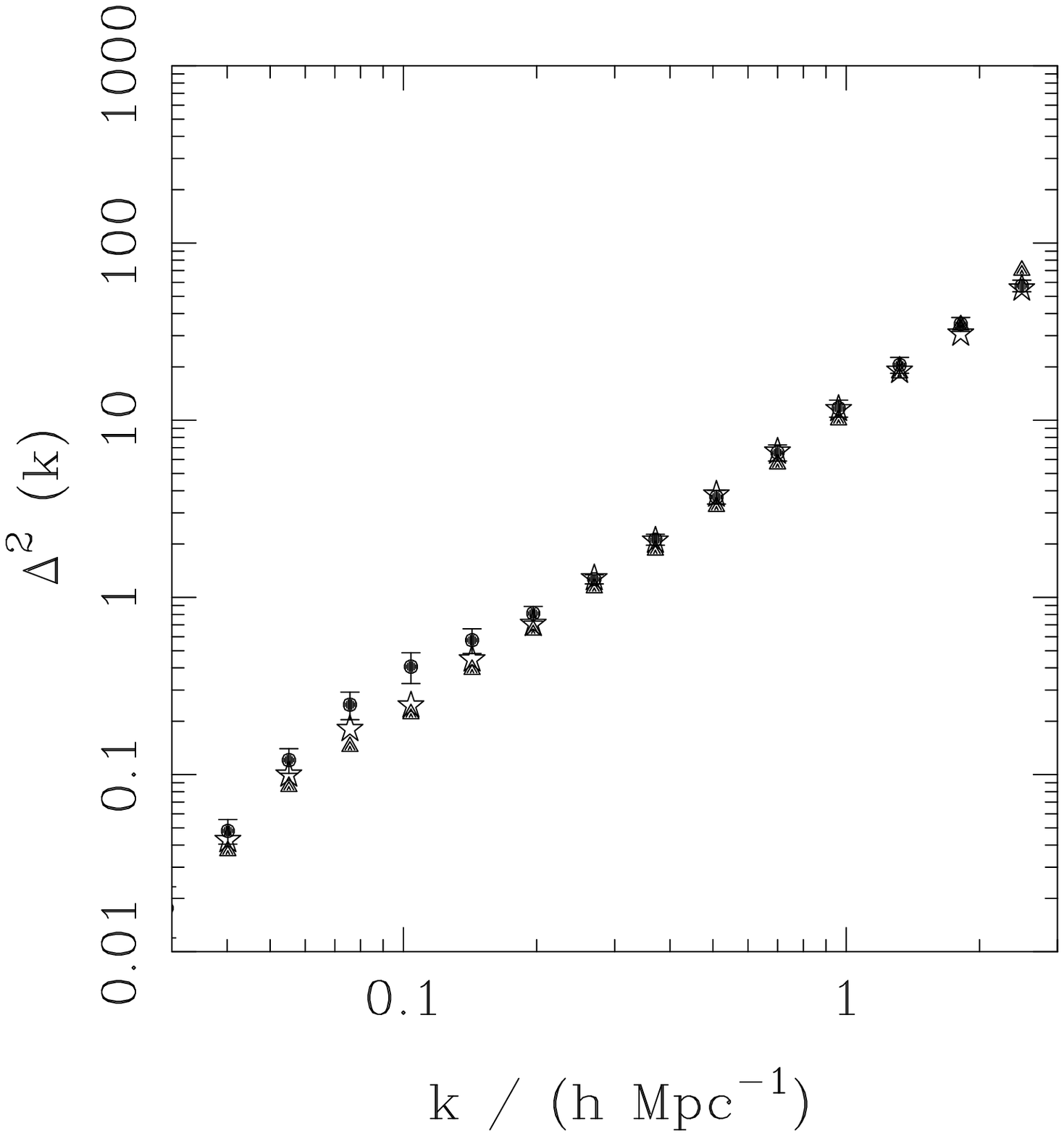,angle=0,width=8.5cm}
\caption{A comparison between the APM data (filled circles)
and two power law biased CDM models:
the stars show $\Gamma^*\simeq 0.17, \; \sigma_8=1.0, \; B=0.85$,
and the triangles show $\Gamma^*\simeq 0.17, \; \sigma_8=0.55, \; B=1.3$.}
\efi

The best-fitting values of $B$ for the models of Fig. 10 are plotted in
Fig. 12.  
The contour lines (for $B \ga 0.8$) are very close to being lines of 
constant $\sigma_8$.   Since $\sigma_8$ is related roughly to a scale
\mbox{$k \sim 0.3 \hompc$}, which is close to the asymptotic region
where $b(k)$ is constant, this indicates that the most important effect of
the bias is in setting the normalisation, rather than the detailed
shape, of the galaxy power spectrum: this bears out the results given
in equation (21). Note that Fig. 10 shows that $\chi^2$ decreases slightly
as $\sigma_8$ falls below 0.4 for $\Gamma^*=0.15-0.20$, and,  from
Fig. 12, the best fit $B$ values $B \sim 1.5$ are much higher than
those near the global $\chi^2$ minimum. This indicates how, given the
linear power spectrum of a particular mass model,
there are two ways to try to achieve the steepening required to match the
APM power spectrum: use a high bias on a moderately evolved mass
distribution, or evolve the mass distribution much more and use
anti-bias ($B<1$) to pull down the power to the desired level.

\bfi
\epsfig{file=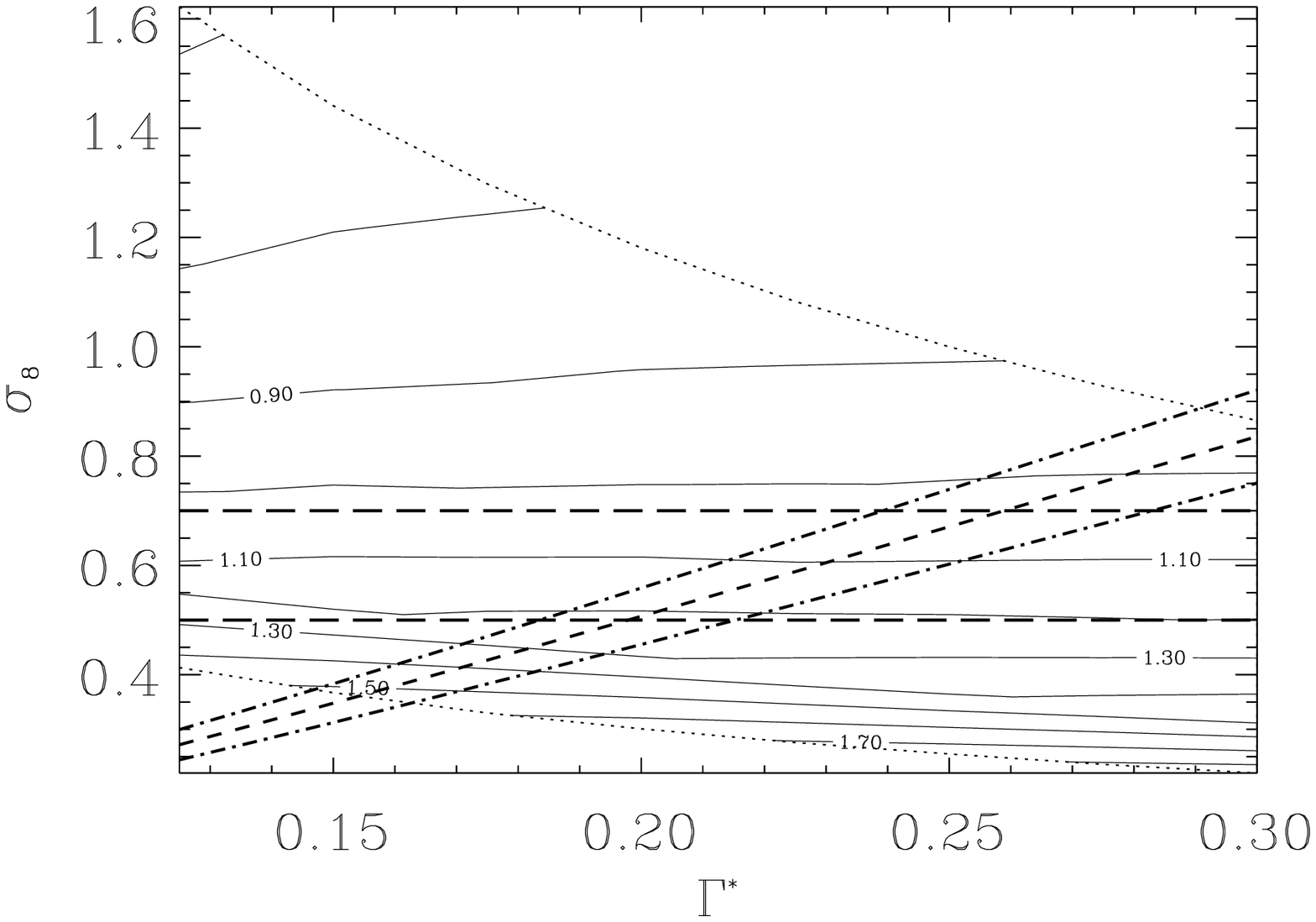,angle=0,width=8.5cm}
\caption{The best-fitting value of $B$ in power law bias models, as a
function of $(\Gamma^*,\sigma_8)$. 
The horizontal lines show the allowed range of $\sigma_8$ from the
cluster abundance constraint.
The diagonal lines show the constraint required to
satisfy the {\it COBE\/} measurement for the case of no tilt.
}
\efi

\subsubsection{Cen-Ostriker bias}

For the remainder of our analysis we restrict our attention to the region of 
the $(\Gamma^*,\sigma_8)$ plane which satisfies both our normalisation
criteria. 
The form of the $\chi^2(C_1,C_2)$ surface for these mass models has a
very characteristic form, as illustrated by Fig. 13, which shows it 
for one such model, with $(\Gamma^*=0.25$ and $\sigma_8=0.64)$. In all
cases the $\chi^2$ contours form a wedge-shaped valley in the $(C_1,C_2)$ 
plane, the gross features of which are
readily explained. The introduction of the $C_2$ term into the bias
formula allows these mass models to improve agreement with the APM
power spectrum by maintaining their large-scale power  (by
increasing $C_1$ above the value of the best-fit $B$ value) in the
face of its suppression by the negative $C_2$ value which is needed so
as not to produce too much small-scale power.
The floor of the valley in the $\chi^2$
surface gently rises with increasing $C_1$, but we need not concern
ourselves with whether or not models matching the APM power spectrum
are to be found at large $C_1$, since the increasingly negative value
of $C_2$ required to make that match very rapidly reduces the predicted
cluster $M/L$ ratio to very low values.  
In fact, acceptable $\chi^2$ values are rarely found above $C_1=2$,
and the negative $C_2$ values required by acceptable models means 
that their mass-to-light ratios fall short by factors of several
hundred from that required to reconcile observed cluster $M/L$
values with $\Omega=1$. We therefore exclude all 
CDM-like models biased according to the Cen-Ostriker  formula.

\bfi
\epsfig{file=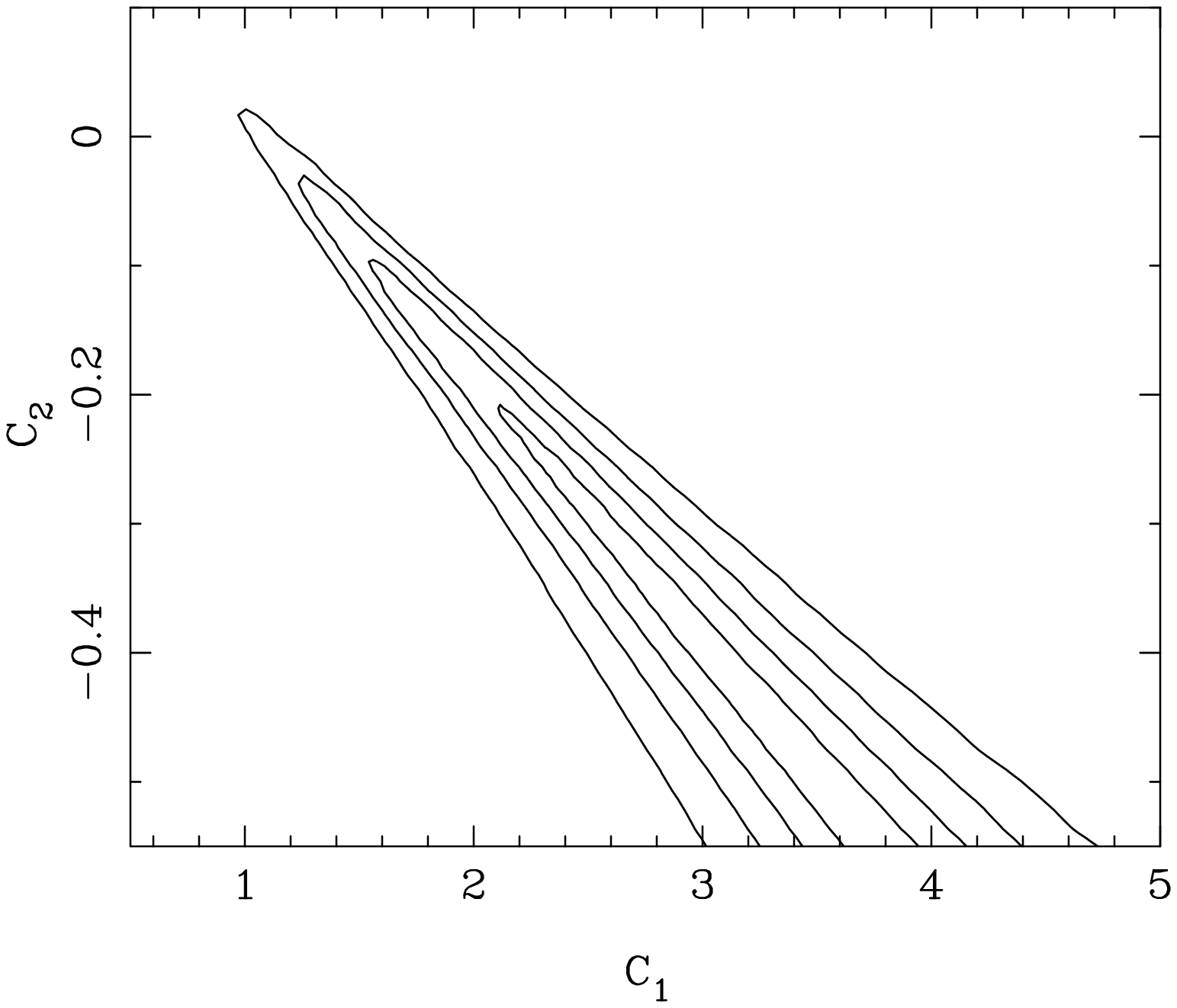,angle=0,width=8.5cm}
\caption{Contours of $\chi^2$ as
a function of the CO parameters $C_1$ \& $C_2$
in the comparison between biased CDM power spectra
and the APM data.
The contours are at $\Delta(\log_{\rm 10}\chi^2)=0.3$ above $\chi^2=1$.
}
\efi

\subsubsection{Censored models}

In Fig. 14 we plot  $\chi^2(B,f_{\rm M})$ for a
particular allowed mass model ($\Gamma^*=0.25,\sigma_8=0.64$
again). The basic form of this curve is, again, characteristic and
easy to understand: as more
and more weakly-clustered mass is censored, a lower value of $B$ is
required to match the APM power spectrum to a given accuracy. Typically, the
best-fit censored power law model has $f_{\rm M} \simeq 0.35-0.45$ and a
value of $B$ lower than that for the best pure power law bias
model. Those censored power law 
bias models that give a good fit to the APM power spectrum may be
excluded on $M/L$ ratio grounds, as is clear from Fig. 14, which shows
that increasing $f_{\rm M}$ only leads to lower values of $B$: e.g. censoring 
half the mass increases the mass-to-light ratio by a factor of two, but
the low value of $B$ reduces it by a very much larger factor.

\bfi
\epsfig{file=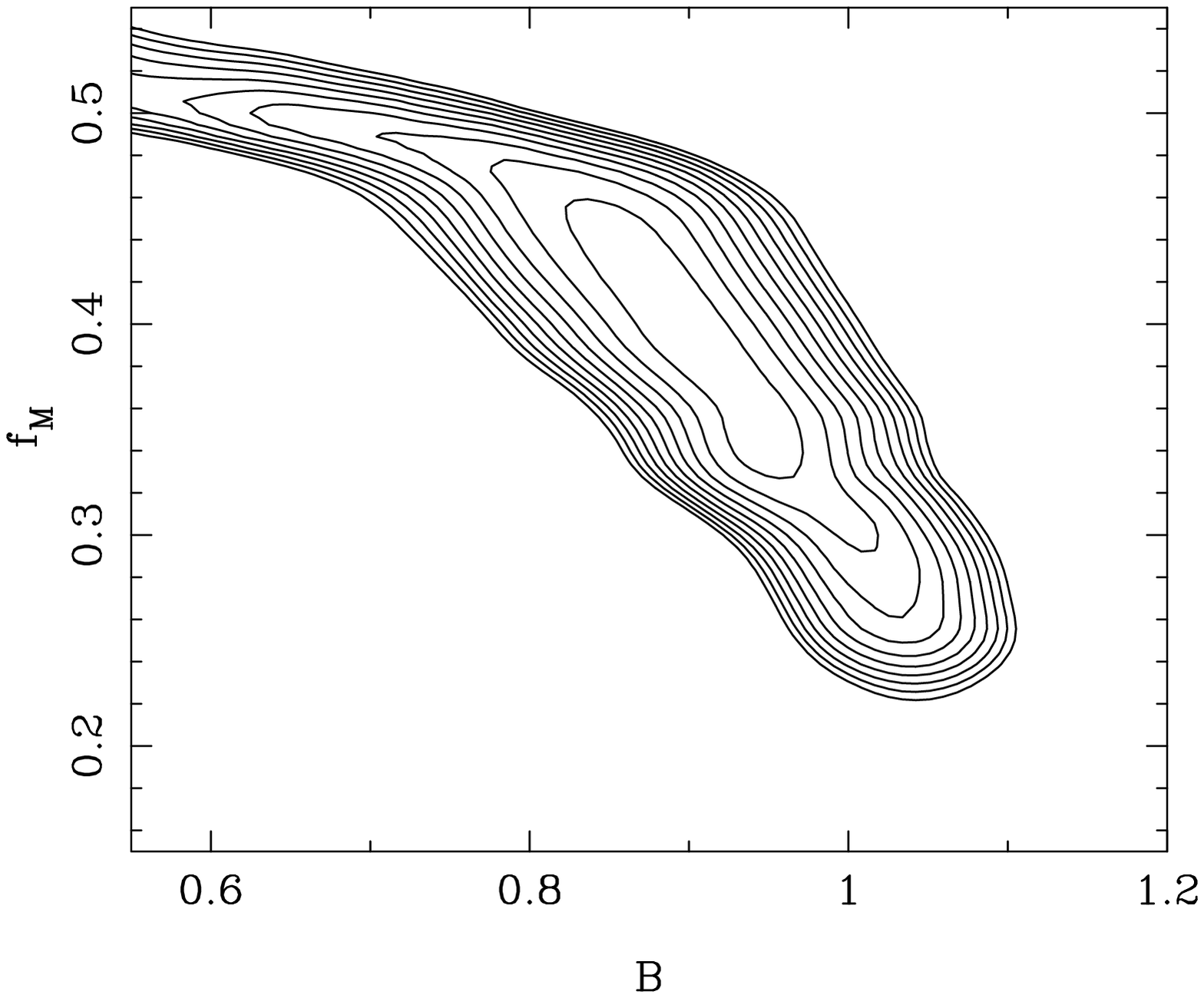,angle=0,width=8.5cm}
\caption{Contours of $\chi^2$ 
for the bias model with power law bias and censoring, as
a function of the parameters $B$ and $f_m$,
in the comparison between biased CDM power spectra
and the APM data.
The contours are $\Delta \chi^2=15$, with the inner contour being $\chi^2=30.$
}
\efi

\subsection{Comparison of bias prescriptions}

An enlightening comparison of the effects of the three bias
prescriptions is provided by Figs. 15 and 16. In Fig. 15, the
filled circles show the APM galaxy power spectrum, while the stars
of David show the mass power spectrum for a model with
$\Gamma^*=0.25$ and $\sigma_8=0.64$. Clearly, to match
the APM power spectrum, this mass model needs a significant bias on 
large scales, below the inflection in the APM power spectrum. However,
the required level of bias falls with increasing wavenumber, and
there is little room for bias on the smallest scales. The best fit
power law bias model yields $\chi^2=96.2$ at $B=1.07$, and its
power spectrum is shown by the stars in Fig. 15, where it is clear
that only a modest power law bias produces too much small-scale power,
while falling short of the desired clustering strength on large
scales. The freedom to censor mass produces the power spectrum given
by the triangles in Fig. 15, which has $B=0.93$ and $f_{\rm M}=0.38$
(giving $\chi^2=62.3)$, while the best fit Cen-Ostriker model is
$C_1=2.0$, $C_2=-0.19$, (crosses), which yields $\chi^2=27.1$. The
different effects of these three models are better shown by their 
$b(k)$ curves, which are plotted in Fig. 16. Censoring almost
40 per cent of the mass produces the required increase in large-scale
power, but $B<1$ is required to restrict the level of small-scale
clustering, which still exceeds that of APM galaxies. The great
freedom afforded by the introduction of the $C_2$ parameter is
shown by the much larger $\chi^2$ reduction it allows, as well as
by the shape of the $b(k)$ curve it produces in Fig. 16: a much
better fit to the APM power spectrum is obtained, as equation (10)
allows a substantial large-scale bias at the same time as keeping
the small-scale clustering power pinned, rising only slighty above the
level of that found in the mass distribution.
The problem with this model, of course, is that it fails to
provide a sufficiently large boost to $M/L$ on scales of clusters.
There appears to be no way in which we can achieve the necessary
$b(k)$ without violating this constraint.

\begin{figure}
\epsfig{file=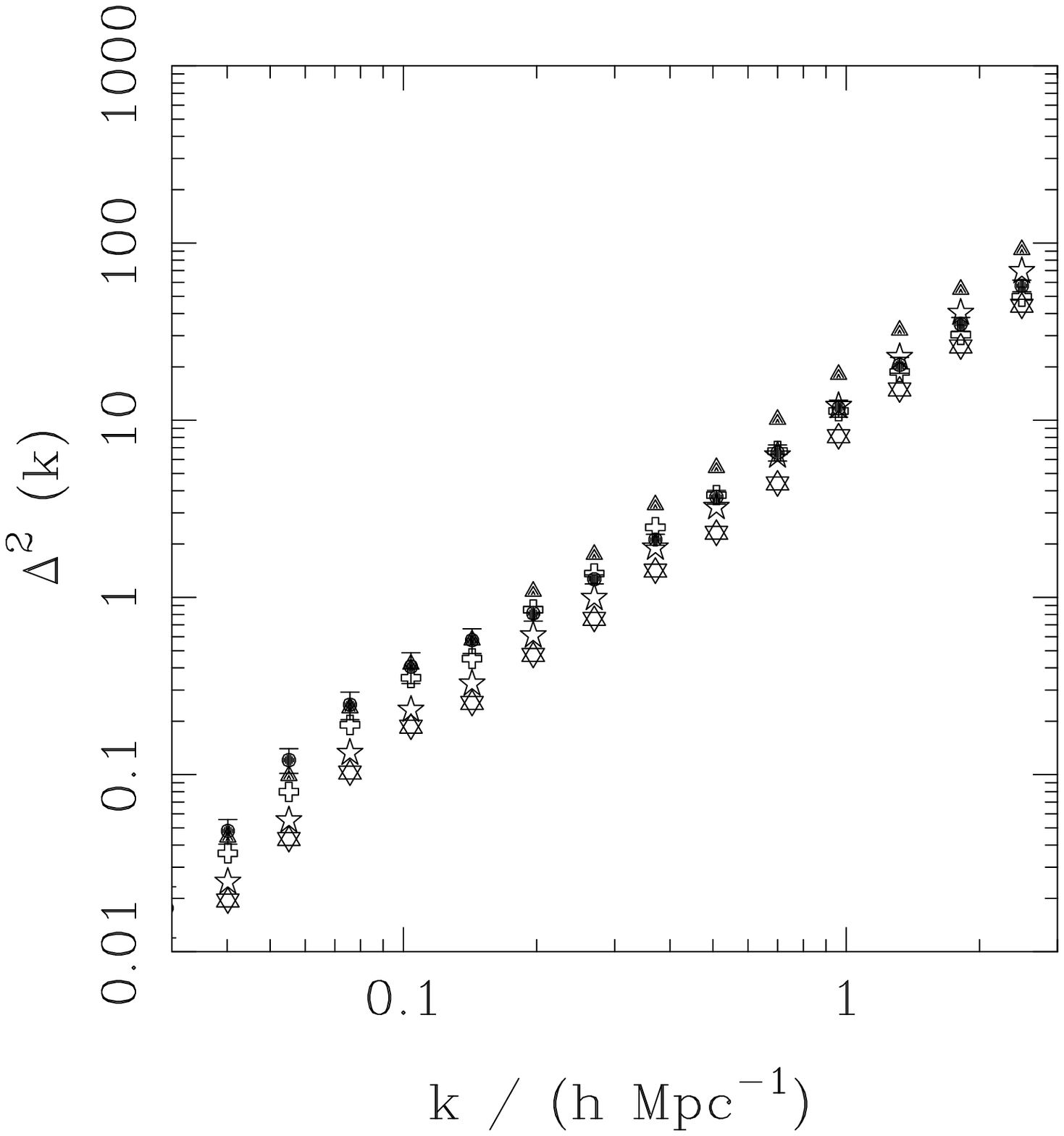,angle=0,width=8.5cm}
\caption{Biasing a mass model to fit the APM power spectrum. The mass
power spectrum for a model with $\Gamma^*=0.25$ and $\sigma_8=0.64$
is shown by the stars of David, while the filled circles plot the
APM power spectrum. The best fit models from the three bias
prescriptions are shown as follows: (a) power law bias with $B=1.07$
(stars); (b) censored power law bias with $B=0.93$ and $f_{\rm
M}=0.38$ (triangles); and (c) Cen-Ostriker bias with $C_1=2.0$ and
$C_2=-0.19$ (crosses).}
\end{figure}

\begin{figure}
\epsfig{file=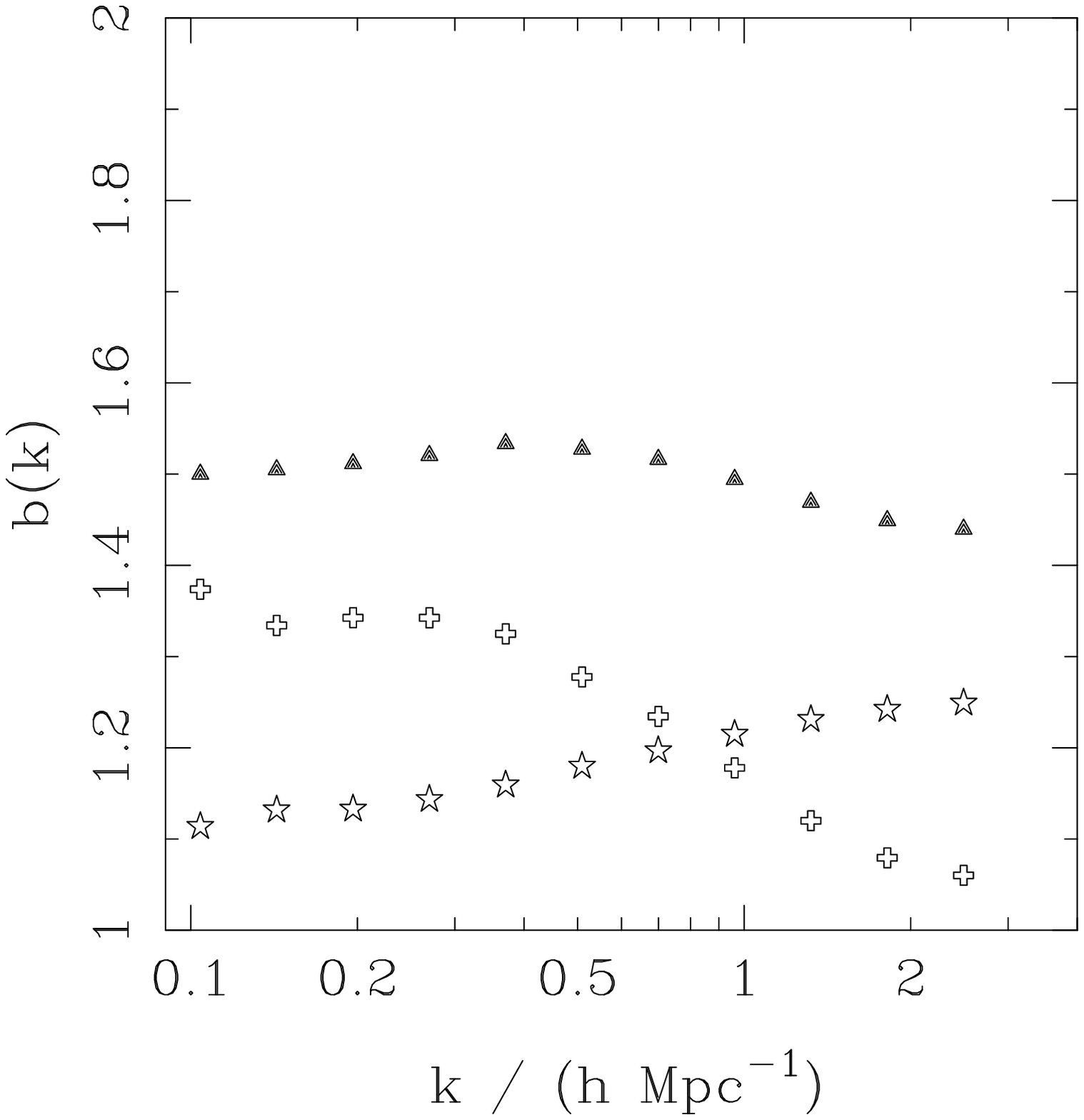,angle=0,width=8.5cm}
\caption{The scale dependence of the bias in the three galaxy power
spectra shown in Fig. 15: (a) power law bias with $B=1.07$
(stars); (b) censored power law bias with $B=0.93$ and $f_{\rm
M}=0.38$ (triangles); and (c) Cen-Ostriker bias with $C_1=2.0$ and
$C_2=-0.19$ (crosses).}
\end{figure}

\section{DISCUSSION AND CONCLUSIONS}

We have investigated the effect of local biasing on the clustering
statistics of galaxies.   In general, we find that local biasing 
gives rise to a scale-dependent bias, so the shape of galaxy power spectrum
no longer directly reflects the underlying mass distribution. However,
the change in $b(k)$ with scale is often rather modest:
if $b_\infty \simeq 1.5$, as suggested
by observations and {\em COBE} normalisation, then the
models we have investigated would only predict
at most $b(k) \simeq 3$ at $k \simeq 1  \hompc$.
The scale dependence of the bias  appears to be monotonic, but $b(k)$
does not decrease with scale in all cases, showing how the
inequalities of Coles (1993) break down once the density field is 
sufficiently non-linear that it is no longer well described by a
Gaussian field, or a local transformation thereof. The
$b(k)$ curves produced by our biasing prescriptions are also quite
smooth, so it would appear that any model capable of producing sharp
changes in galaxy power with respect to that of the mass would have to involve
non-local bias. 

The scale dependence of bias clearly presents 
difficulties in the interpretation of 
observed galaxy clustering, by allowing a wider range of mass power spectra
to map onto the galaxy power spectrum data through the use of a suitably
curved $b(k)$. Interpretation of even these models is simplified
on large scales, however, where the bias tends to a constant value.

We have looked at a range of galaxy clustering models in the context of an
Einstein - de Sitter universe, to see if any of these models can 
simultaneously fit the APM power spectrum and the observed mass-to-light
ratios in clusters of galaxies.   The strong conclusion we reach
is that {\em COBE\/}-normalised CDM-like models with $\sigma_8$ fixed
as required by the observed abundances of clusters do not succeed, given the 
biasing functions we have explored.   
In addition, the models fail to reproduce the required mass-to-light ratios
by large factors, so if any successful bias function were found, it would 
probably be rather pathological.

There are several ways to improve the agreement of the models with observation.
One is to lower the density below $\Omega=1$, thus reducing or removing the
constraints on our models in the high-density regions.  The lower $\Omega$ 
has the effect of increasing the required $\sigma_8$ from cluster
abundances, but the non-linear effects on the mass spectrum depend
on $\Omega$ only on scales smaller than those we have considered here
(Peacock \& Dodds 1996).
The second is to change the shape of the underlying
power spectrum to one with a sharper break than CDM, such as mixed dark matter
(e.g. Taylor \& Rowan-Robinson 1992;  van Dalen \& Schaeffer 1992;
Klypin et al. 1993). 
This latter possibility is
the only existing model which can explain the full inflection
in the APM power spectrum around $k\simeq 0.1 \hompc$, since
CDM-like models produce a much smoother variation
at this point.

The conclusions we draw from our results are thus as follows:

\begin{itemize}

	\begin{enumerate}
	
		\item No linearly-biased CDM-like model can reproduce
the APM power spectrum: a non-linear $b(k)$ curve is required if the
cosmological fluctuations spectrum is one of the $\Gamma^*$ models
given by equations (12) and (13). Analyses (such as that of PD94) 
which do allow a linearly-biased $\Gamma^*$ model can do so only
because they ignore clustering data on non-linear scales, which
greatly weakens their power to reject models.

		\item Non-linear bias prescriptions generically
produce a scale-dependent bias, but it is difficult to produce a
strongly curved $b(k)$ relation with Eulerian bias, so a confirmation of the
APM inflection would be strong evidence for some feature
in the primordial linear spectrum at that scale or for non-local bias
in galaxy clustering.

		\item Analytical models based on a lognormal mass
distribution are useful heuristics for gauging the rough level of the
large-scale bias produced by a given prescription, but they are not
sufficiently accurate to replace numerical simulations.

		\item No Einstein -- de Sitter CDM-like mass model can
be biased, using the
range of bias prescriptions studied here, to match the APM power
spectrum if it is required to be {\em COBE\/}-normalised and account for
the abundances and $M/L$ values of clusters of galaxies: if these
conditions are to be satisfied using the bias models considered here,
then the mass power spectrum must be
flatter than CDM-like models on small scales.  

		\end{enumerate}

\end{itemize}

\section*{ACKNOWLEDGEMENTS} 

RGM acknowledges support from PPARC rolling
grants at QMW and Imperial College. We thank Ed Bertschinger and
Jim Gelb for numerical simulation data used at an early stage of
this project, John Huchra for the use of the ZCAT redshift
compilation and an anonymous referee for comments that clarified
the presentation of several points in this paper.

\section*{REFERENCES}

\myref{Babul A., White S.D.M., 1991, MNRAS, 253, 31{\sc p}}
\myref{Bardeen J.M., Bond J.R., Kaiser N., Szalay A.S., 1986, ApJ, 
304, 15 (BBKS)}
\myref{Baugh C.M., Efstathiou G., 1993, MNRAS, 265, 145} 
\myref{Catelan P., Coles P., Matarrese S., Moscardini L., 1994, MNRAS,
268, 966} 
\myref{Cen R.Y., Ostriker J.P., 1992, ApJ, 399, L13 (CO)}
\myref{Cole S., Kaiser N., 1989, MNRAS, 237, 1127}
\myref{Coles P., 1993, MNRAS, 262, 1065}
\myref{Coles P., Jones B., 1991, MNRAS, 248, 1}
\myref{Couchman H.M.P., 1991, ApJ, 368, L23}
\myref{Davis M., Efstathuiou G., Frenk C.S., White S.D.M., 1985, ApJ,
292, 371}
\myref{Dekel A., Rees M.J., 1987, Nat, 326, 455}
\myref{Efstathiou G., Bond J.R., White S.D.M., 1992, MNRAS, 258, 1{\sc p}}
\myref{Hamilton A.J.S., Kumar P., Lu E., Matthews A., 1991, ApJ, 374, L1}
\myref{Huchra J., Geller M., Clemens C., Tokarz S., Michel A, 1992, Bull. C.D.S. 41, 31}
\myref{Kaiser N., 1984, ApJ, 284, L9}
\myref{Kauffmann G., Nusser A., Steinmetz M., 1997, MNRAS, 286, 795}
\myref{Katz N., Quinn T., Gelb J.M., 1993, MNRAS, 265, 689}
\myref{Klypin A., Holtzman J., Primack J., Reg\H os E., 1993, ApJ, 416, 1}
\myref{Little B., Weinberg D.H., 1994, MNRAS, 267, 605}
\myref{Mann R.G., Heavens A.F., Peacock J.A., 1993, MNRAS, 263, 798}
\myref{Matarrese S., Coles P., Lucchin F., Moscardini L., 1997, MNRAS,
286, 115}
\myref{Matarrese S., Verde L., Heavens A.F., 1997, MNRAS, in press
(astro-ph/9706059)}
\myref{Mo H.J., Jing Y.P., White S.D.M., 1996, MNRAS, 282, 1096}
\myref{Mo H.J., White S.D.M., 1996, MNRAS, 282, 347}
\myref{Peacock J.A., 1997, MNRAS, 284, 885}
\myref{Peacock J.A., Dodds S.J., 1994, MNRAS, 267, 1020 (PD94)}
\myref{Peacock J.A., Dodds S.J., 1996, MNRAS, 280, L19}
\myref{Sugiyama N., 1995, ApJS, 100, 281}
\myref{Taylor A.N., Rowan-Robinson M., 1992, Nat, 359, 396}
\myref{Tegmark M., 1997, ApJ, 464 L35} 
\myref{van Dalen A., Schaefer R.K., 1992, ApJ, 398, 33}
\myref{White S.D.M., Efstathiou G., Frenk C.S., 1993, MNRAS, 262, 1023}

\appendix
\setcounter{equation}{0}
\renewcommand\thesection{\Alph{section}}
\renewcommand\theequation{\thesection\arabic{equation}}

\section{BIASED LOGNORMAL DENSITY FIELDS}

The lognormal density field is a convenient analytical model 
in which the effects of various simple Eulerian density
transformations can be calculated quite simply.
The non-linear lognormal density field is produced
by taking the exponential of a Gaussian density field:
\beq
1+\delta_{\rm LN}=\exp\left(\delta_{\rm G}- \sigma^2/2\right).
\eeq
The variance of the Gaussian field is $\sigma^2$ and the second term in the
exponential is a normalising factor, so that $\langle \delta_{\rm LN}
\rangle=0$.
Although this transformation resembles a form of bias,
and has been discussed in that context, it is important to be
clear that this is not what is envisaged here. The lognormal
field is a model for the mass density field: it is the simplest
extension of the Gaussian random field which is physical in the
sense of forbidding negative densities. The lognormal field
differs from its generating Gaussian only when $\sigma$ is
significantly larger than zero, so the lognormal transformation should
be thought
of as a means of simulating the effects of non-linear gravitational
instability.

\subsection{Power law bias}

The simplest modification of a density field is a power law
\beq
\rho \propto \rho_{\rm LN}^B \propto \exp(B\delta_{\rm G}).
\eeq
The effects of this modification are easily deduced, since Coles \&
Jones (1991) showed that 
\beq
1+\xi_{\rm LN}=\exp(\xi_{\rm G}).
\eeq
The power law model is just the original lognormal model with the
generating Gaussian scaled by a factor $B$, and so
\beq
1+\xi_{\rm PLN}=\exp (B^2 \xi_{\rm G}) = (1+\xi_{\rm LN})^{B^2}.
\eeq
Interestingly, this modification does not increase the skewness of the
density field. We define the skewness parameter (not the skewness
itself) as
\beq
S \equiv \frac {\langle \delta^3 \rangle}{\left( \langle \delta^2
\rangle \right)^2}.
\eeq
Integrating over the Gaussian field gives
\beq
S=2+ {\rm e}^{B^2 \sigma^2}.
\eeq
Since the observed density rms in the linear regime is just $B\sigma$,
this says that the observed skewness-variance relation is unaltered
by bias, which masquerades as extra non-linearity in the mass field.
This degeneracy between bias and evolution is a feature of the lognormal
model.  Genuine gravitational evolution leads to different shaped 
structures (flattened on first collapse), and bias and evolution can 
be separated by studying shape-dependent statistics such as the bispectrum
or 3-point correlation function (e.g. Matarrese, Verde \& Heavens 1997).

\subsection{Censoring bias}

A power law provides a significant boost in density at large
over-densities, but it is also plausible that the galaxy density
may suffer abrupt truncation at small over-densities, i.e. in the
voids.

In the case of no other bias, this is the model studied by Catelan et
al. (1994). The method of evaluation is the same in all such cases:
write $\delta^{\prime}=f(\delta)$, where it is
assumed that the function $f$ is normalised to force $\langle
\delta^{\prime} \rangle=0$, and use 
\beq
1+\xi'=\langle(1+\delta^{\prime}_1)(1+\delta^{\prime}_2) \rangle.
\eeq
The correlation function is then
\beq
1+\xi^{\prime}=\int_{-\infty}^{\infty} \int_{\infty}^{\infty}
(1+f_1)(1+f_2) \; P(\delta_1,\delta_2) \; {\rm d}\delta_1 {\rm
d}\delta_2,
\eeq
where the Gaussian joint distribution involves the correlation
coefficient $r \equiv \xi/\sigma^2$.
For the case of truncated power law bias, the modified density field is
\beq
1+\delta' = { \exp[B\delta_{\rm G} -B^2\sigma^2/2] \over
 0.5 \left[ 1- {\rm erf}\left({\nu-B\sigma\over\sqrt{2}}\right)\right]},
\eeq
where the threshold is specified in terms of its value in the Gaussian
field that generates the mass: $\delta'=-1$ for $\delta_{\rm G}<\nu\sigma$.

For the case of censored power law bias, the correlation function reduces 
to the 1D integral
\begin{eqnarray}
1 & + & \xi^{\prime} =  \sqrt{2/\pi} \, \frac {{\rm e}^{-B^2\sigma^2
(1+r^2)/2}} {\left[1-{\rm erf}{(\nu-B\sigma)/\sqrt{2}} \right]^2} \; \nonumber \\
& & \nonumber \\
& \times & \!\!\!\!\!\!\int_\nu^\infty \!\!\!\!\!\!\!
{\rm e}^{xB\sigma(1+r)-x^2/2} \!\! 
\left[1-\!{\rm erf}\!\left( \frac{\scriptstyle \nu-B\sigma+r^2B\sigma-rx}{
\scriptstyle \sqrt{2(1-r^2)}}
\right) \right]  {\rm d}x.
\end{eqnarray}

In the $\xi \ll 1$ regime, this reduces to $\xi'=b_{\rm LN}^2 r \sigma^2$, with
the linear bias parameter
\beq
b_{\rm LN}= B + \frac {\sqrt{2/\pi} \; {\rm e}^{-(\nu-B\sigma)^2/2}}
{\sigma \; \left[1 -{\rm erf} \left( \frac {\nu-B\sigma} {\sqrt{2}} \right) \,
\right].}
\eeq
which reduces to the result of Catelan et al. (1994) for $B=1$, once
allowance is made for a spurious factor of $1/\sqrt{2}$ in their
equation (19).

The results for the case of censoring (with or without power law
bias) are hardly transparent, and it may be helpful to illustrate
them in one particular case. Consider the following model 
for the galaxy correlation function, shown in Fig. A1:
\beq
\xi = [\xi_{\rm max}^{-1} + (r/r_0)^\gamma]^{-1}.
\eeq
For the case of pure power law bias, it is simple
to recover the corresponding mass correlation function,
and to obtain from this the correlation function for
the Gaussian process which would generate the mass
field as a lognormal process. We shall consider the
specific case $B=1.23$.
We also take $\xi_{\rm max}=10^4$,
giving $\sigma=2.47$, although this is not critical.

We now compare the action of censoring bias on the
same mass field. To achieve the same linear bias of
1.23 requires a threshold $\nu=2.1$, corresponding to
censoring 98 per cent of the volume and 36 per cent of the mass.
As a perhaps more realistic compromise model, we consider
censoring and power law bias in combination.
A threshold of $\nu=1.3$ censors 90 per cent of the volume
and 12 per cent of the mass, and requires $B=1.18$ for the
same linear bias. These alternatives are compared in Figure
A1. We see that censoring alone steepens the correlation function,
but not to the same extent as power law bias with the
same large-scale bias. An excellent fit (not shown, as it
is hard to see any deviation) is obtained with the two parameter
formula
\beq
1+\xi' = (1+b_1 \xi)^{b_2},
\eeq
which is of course a perfect fit in the power law only case.
For censoring only, $(b_1,b_2)=(1.35, 1.12)$ is needed,
changing to $(1.09,1.39)$ in the combination case.

Although this is only one example, the general behaviour
follows this trend: for a given linear bias, censoring is
less effective at steepening the correlation function than
is power law bias. To increase the small-scale correlations
by more than a factor $\sim 3$ requires very severe
thresholding in which galaxies only form in the very densest
of environments.

\bfi
\epsfig{file=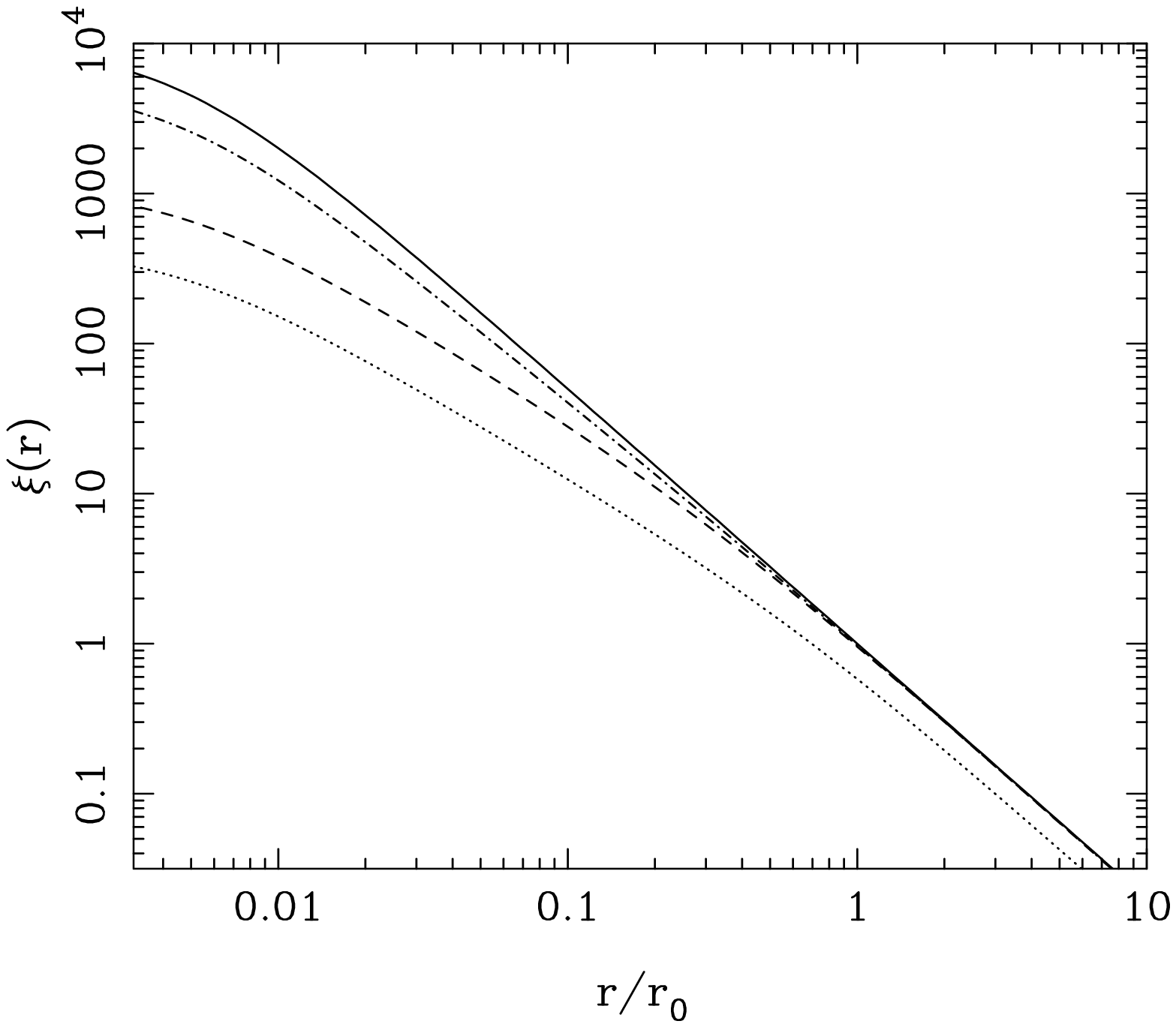,angle=0,width=8.5cm}
\caption{
A plot of various correlation functions. The dotted line
shows $\xi$ for the mass, set so that pure power law bias
with $B=1.23$ produces $\xi'$ which is a reasonable match
to observed galaxy clustering ($\gamma=1.7$, with clustering
extending up to $\xi=10^4$). The three models shown all
have the same linear $b_\infty=1.23$. In order of
increasing $\xi'$, they are (1) censoring only, rejecting
36 per cent of the mass; (2) censoring 12 per cent of the
mass and power law bias with $B=1.18$; (3) B=1.23 with
no censoring. }

\efi

\subsection{Cen-Ostriker bias}

A final example which is amenable to analytical treatment is
the case considered by Cen \& Ostriker (1992), where there is a
quadratic relationship between the logarithms of the light and
mass densities:
\beq
\ln(\rho_{\rm g}/\bar{\rho}_{\rm g})= C_0 + C_1\ln(\rho_{\rm
m}/\bar{\rho}_{\rm m}) + C_2 [\ln(\rho_{\rm m}/\bar{\rho}_{\rm m})]^2.
\eeq
Note that for $C_2=0$, this is just the earlier power law model, and that
the maximum value for $C_2$ is $1/2\sigma^2$, as the mean density
diverges if that limited is exceeded.
If the density field is lognormal, then the properly 
normalized version of this becomes
\begin{eqnarray}
1&+&\delta'=\sqrt{1-2C_2\sigma^2}\, \nonumber\\
& & \nonumber \\
&\times&\!\!\!\!\!\exp\!\!\left[(C_1-C_2\sigma^2)\delta_G + \!C_2\delta^2_G -
\!\!{(C_1-C_2\sigma^2)^2\sigma^2\over 2(1-2C_2\sigma^2)} \right].
\end{eqnarray}
The integral for $\xi'$ can be performed in this case, giving
\begin{eqnarray}
1&+&\xi'={(1-2C_2\sigma^2)\sqrt{1-r^2} \over \sqrt{[1-2C_2\sigma^2(1-r^2)]^2-r^2}} \nonumber \\
& & \nonumber \\
&\times&\exp\left[{(C_1-C_2\sigma^2)^2\sigma^2 r \over
1-2C_2\sigma^2(2+r)+4C_2^2\sigma^4(1+r)}\right].
\end{eqnarray}
This has the linear bias
\beq
b_{\rm LN}= { C_1-C_2\sigma^2 \over 1-2C_2\sigma^2 },
\eeq
which gives the power law case when $C_2=0$. However, we also see
that the bias is weakened when $C_2$ is negative, so that the
light density is suppressed at extreme overdensity: the linear
bias is unity when
\beq
C_2=C_{\rm2, crit}=-(C_1-1)/\sigma^2.
\eeq

\bsp

\end{document}